\documentclass[11pt,a4paper]{article}

\usepackage{epsfig}
\usepackage{amsmath}
\usepackage{amssymb}
\usepackage{verbatim}
\usepackage{bm}
\usepackage{dsfont}
\usepackage[footnotesize]{caption}
\usepackage{subfigure}
\usepackage{cite}
\usepackage{textcomp}
\usepackage{calc}
\usepackage{geometry}
\usepackage{bbm}
\usepackage{xcolor}
\usepackage{multirow}
\usepackage{color}
\usepackage{float}
\usepackage{bbold}
\usepackage{ulem}

\usepackage{hyperref}
\hypersetup{colorlinks=true,urlcolor=blue,linkcolor=red,citecolor=green!60!black}

\renewcommand{\baselinestretch}{1.3}

\begin{document}


\thispagestyle{empty}

\noindent July 2017
\hfill 

\vskip 1.5cm

\begin{center}
{\LARGE\bf Baryon Asymmetry and Gravitational Waves\\\smallskip from Pseudoscalar Inflation}

\vskip 2cm

\renewcommand*{\thefootnote}{\fnsymbol{footnote}}

{\large
Daniel Jim\'enez,$^{a,\,\hspace{-0.25mm}}$%
\footnote{daniel.jimenez@mpi-hd.mpg.de}
Kohei Kamada,$^{b,\,\hspace{-0.25mm}}$%
\footnote{kohei.kamada@asu.edu}
Kai~Schmitz,$^{a,\,\hspace{-0.25mm}}$%
\footnote{kai.schmitz@mpi-hd.mpg.de}
and Xun-Jie~Xu\,$^{a,\,\hspace{-0.25mm}}$%
\footnote{xunjie.xu@gmail.com}}\\[3mm]
{\it{
$^{a}$ Max-Planck-Institut f\"ur Kernphysik (MPIK), 69117 Heidelberg, Germany\\
$^{b}$ School of Earth and Space Exploration, Arizona State University, Tempe, AZ 85287, USA}}

\end{center}

\vskip 1cm

\renewcommand*{\thefootnote}{\arabic{footnote}}
\setcounter{footnote}{0}


\begin{abstract}


In models of inflation driven by an axion-like pseudoscalar field, the inflaton, $a$,
may couple to the standard model hypercharge via a Chern-Simons-type interaction,
$\mathcal{L} \supset a/\left(4\Lambda\right) F\tilde{F}$.
This coupling results in explosive gauge field production during inflation,
especially at its last stage, which has interesting phenomenological consequences:
For one thing, the primordial hypermagnetic field is maximally helical.
It is thus capable of sourcing the generation of nonzero baryon number,
via the standard model chiral anomaly, around the time of electroweak symmetry breaking.
For another thing, the gauge field production during
inflation feeds back into the primordial tensor power
spectrum, leaving an imprint in the stochastic background
of gravitational waves (GWs).
In this paper, we focus on the correlation between these two phenomena.
Working in the approximation of instant reheating, we
(1) update the investigation of baryogenesis
via hypermagnetic fields from pseudoscalar inflation and
(2) examine the corresponding implications for the GW spectrum.
We find that successful baryogenesis requires a suppression scale $\Lambda$ of around
$\Lambda \sim 3 \times 10^{17}\,\textrm{GeV}$,
which corresponds to a relatively weakly coupled axion.
The gauge field production at the end of inflation is then
typically accompanied by a peak in the GW spectrum at frequencies in the MHz range or above.
The detection of such a peak is out of reach of present-day technology;
but in the future, it may serve as a smoking-gun signal for baryogenesis
from pseudoscalar inflation.
Conversely, models that do yield an observable GW signal suffer from the overproduction
of baryon number, unless the reheating temperature is lower than the electroweak scale.


\end{abstract}


\newpage

{\hypersetup{linkcolor=black}\renewcommand{\baselinestretch}{1} \tableofcontents}


\section{Introduction}
\label{sec:introduction}


In this paper, we are going to study general models of pseudoscalar inflation
and their implications for the present-day spectrum of gravitational waves
as well as for baryogenesis via primordial hypermagnetic fields around the time
of electroweak symmetry breaking (EWSB).
In the following, we will review the status of gravitational waves from
pseudoscalar inflation in Sec.~\ref{subsec:intro_GWs} and baryogenesis
after primordial magnetogenesis in Sec.~\ref{subsec:intro_BAU}.
Readers familiar with both subjects may directly skip to Sec.~\ref{subsec:synopsis},
where we outline the philosophy behind our analysis.


\subsection{Gravitational waves from an anomalous inflaton coupling to gauge fields}
\label{subsec:intro_GWs}


The celebrated detection of gravitational waves (GWs) from a binary black hole merger
by the LIGO/Virgo collaboration~\cite{Abbott:2016blz}
(see also \cite{Abbott:2016nmj,Abbott:2017vtc})
has literally ringed in the era of gravitational-wave astronomy.
In the near future, GW experiments will develop
into standard observational tools, allowing us to routinely
observe\,---\,or better: listen to\,---\,a variety of astrophysical phenomena.
But also from the perspective of particle physics and cosmology,
the observation of GWs bears a huge potential.
In particular, the stochastic background of cosmological GWs emitted during
the early universe carries invaluable information on physical processes at
extremely high energies that are hard or even impossible
to access by other means~\cite{Maggiore:1999vm}.
Among the different possible mechanisms to generate GWs in the early universe,
a prime example is cosmic
inflation~\cite{Starobinsky:1980te,Guth:1980zm,Sato:1980yn,Linde:1981mu,Albrecht:1982wi},
which unavoidably results in the amplification of the quantum
vacuum fluctuations of
the gravitational field~\cite{Starobinsky:1979ty,Rubakov:1982df,Guzzetti:2016mkm}.
In fact, the direct observation of relic GWs from the epoch of inflation would represent a powerful
probe of the earliest moments of our Universe, complementary to other observables that are
sensitive to the dynamics of inflation, such as, e.g., the temperature
anisotropies of the cosmic microwave background (CMB).
Standard single-field slow-roll inflation, however, predicts a
present-day GW spectral energy density, $\Omega_{\rm GW}^0h^2\left(f\right)$,
that falls short of the current experimental sensitivity by many orders of magnitude,%
\footnote{This estimate depends on the reheating temperature, $T_{\rm rh}$, after inflation.
For $T_{\rm rh} \lesssim \mathcal{O}\left(10^9\right)\,\textrm{GeV}$,
one expects that the GW energy density at frequencies in the
$\mathcal{O}\left(10\cdots100\right)\,\textrm{Hz}$ range is further diluted\,---\,and
hence suppressed w.r.t.\ Eq.~\eqref{eq:OGWr}\,---\,during
the stage of expansion dominated by the coherent oscillations of
the inflaton field~\cite{Nakayama:2008ip,Nakayama:2008wy}.\smallskip}
\begin{align}
\label{eq:OGWr}
\Omega_{\rm GW}^0h^2\left(f\right) \sim 10^{-16} \left(\frac{r}{0.1}\right) \,,
\end{align}
where the primordial tensor-to-scalar ratio $r$ is bounded from above by
the CMB observations of the PLANCK satellite, $r < 0.11$ (95\,\% C.\,L.)~\cite{Ade:2015lrj}.
This estimate needs to be contrasted with the sensitivity
of the Advanced LIGO detector after its first run,
$\Omega_{\rm GW}^0h^2\left(f\right) \sim 10^{-7}$ (95\,\% C.\,L.,
at its most sensitive frequencies,
$f \simeq 20\cdots86\,\textrm{Hz}$)~\cite{TheLIGOScientific:2016dpb}.
This sensitivity is certainly an achievement, but still at least nine orders of
magnitude away from the expected signal from inflation.
Meanwhile, future satellite experiments such as DECIGO~\cite{Seto:2001qf,Kawamura:2006up}
and BBO~\cite{Crowder:2005nr,Harry:2006fi}
promise to reach sensitivities that might suffice to detect GWs from inflation at
$\mathcal{O}\left(0.1\cdots1\right)\,\textrm{Hz}$.
But the realization of these experiments is still uncertain and possibly several decades away.


In view of this situation, one is tempted to ask what mechanism could
potentially enhance the GW signal from inflation.
Here, an interesting possibility\,---\,that has recently received renewed attention in
the literature~\cite{Domcke:2016bkh,Garcia-Bellido:2016dkw}\,---\,is
the boosted production of GWs in models of pseudoscalar
inflation~\cite{Cook:2011hg,Anber:2012du}.
This class of inflationary models is built upon the idea that inflation
is driven by the dynamics of a pseudoscalar pseudo-Nambu-Goldstone boson
(PNGB)~\cite{Freese:1990rb,Adams:1992bn}.%
\footnote{The typical example for a PNGB in physics beyond the standard model
is the QCD axion~\cite{Weinberg:1977ma,Wilczek:1977pj} in the Peccei-Quinn solution
to the strong $CP$ problem~\cite{Peccei:1977hh,Peccei:1977ur}.
PNGBs in extensions of the standard model are, therefore, also often
referred to as axion-like particles or simply axions.
In the following, we will use these terms interchangeably.}
Such fields correspond to pseudoflat directions in field space,
the flatness of which is protected against radiative corrections
by an approximate shift symmetry.
For this reason, axion-like directions
provide a \textit{natural} opportunity to realize slow-roll inflation.
The axionic shift symmetry in models of pseudoscalar inflation may,
in particular, correspond to the nonlinear realization of an approximate, Peccei-Quinn-like
global symmetry $G_{\rm global}$.
Furthermore, if this global symmetry is anomalous under some local
gauge symmetry $G_{\rm gauge}$, the inflaton, $a$, will couple to the field strength
tensor of the corresponding gauge field via an effective Chern-Simons term,
\begin{align}
\mathcal{L}_{\rm eff} \supset - \frac{a}{4\,\Lambda}\, F_{\mu\nu}\tilde{F}^{\mu\nu} \,,
\label{eq:aFF}
\end{align}
where $\tilde{F}^{\mu\nu}$ denotes the dual field strength tensor,
$\tilde{F}^{\mu\nu} = \frac{1}{2}\left(-g\right)^{-1/2} \epsilon^{\mu\nu\sigma\tau}F_{\sigma\tau}$,
and where the suppression scale $\Lambda$ is related to the spontaneous symmetry breaking scale
of $G_{\rm global}$.
Similarly, an effective coupling such as in Eq.~\eqref{eq:aFF} may arise
in compactifications of string theory~\cite{Svrcek:2006yi}.
In heterotic string theory, e.g., the Green-Schwarz mechanism
of anomaly cancellation~\cite{Green:1984sg} gives rise to several
(model-dependent as well as model-independent) axions that couple
to the gauge fields of the theory as in Eq.~\eqref{eq:aFF};
see \cite{Obata:2016oym} for a discussion in the context of pseudoscalar
inflation.


The anomalous coupling in Eq.~\eqref{eq:aFF} now has important implications
for the dynamics of inflation and, eventually, for the present-day spectrum
of GWs.
To see this, one first has to note that the axion-gauge-field coupling
in Eq.~\eqref{eq:aFF} results in the explosive production of gauge quanta
during inflation~\cite{Turner:1987bw,Garretson:1992vt,Anber:2006xt}
(see also \cite{Durrer:2010mq,Meerburg:2012id}).
Depending on the sign of the inflaton velocity, $\dot{a}$,
one of the two helicity modes of the gauge field is exponentially amplified,
such that the resulting field configuration is maximally helical.
This is a direct consequence of the fact that the time-dependent vacuum expectation value
(VEV) of the inflaton field breaks parity invariance during inflation,
$\left<{\dot a}\right> \neq 0$.
As the energy transmitted to the gauge field increases,
the gauge field begins to back-react on the evolution of the inflaton,
effectively contributing another friction term (next to the Hubble friction term)
to its equation of motion~\cite{Anber:2009ua,Barnaby:2011qe,Barnaby:2011vw}.
At the same time, fluctuations in the gauge field configuration result
in additional source terms for the primordial scalar
and tensor perturbations.
Together, these effects have a variety of phenomenological consequences, ranging
from modified predictions for various CMB
observables~\cite{Barnaby:2010vf,Sorbo:2011rz,Linde:2012bt,Obata:2016oym},
over the production of primordial black holes~\cite{Bugaev:2013fya,Linde:2012bt},
to\,---\,and here we finally are\,---\,an enhanced spectrum of
GWs~\cite{Cook:2011hg,Anber:2012du,Domcke:2016bkh,Garcia-Bellido:2016dkw}.


Recently, it has been pointed out that the GW signal from pseudoscalar inflation
may be even amplified to such an extent that it falls into the sensitivity reach of upcoming
GW interferometer experiments~\cite{Domcke:2016bkh,Garcia-Bellido:2016dkw}.
Here, a particularly promising inflation model appears
to be Starobinsky inflation~\cite{Starobinsky:1980te}, which could potentially
lead to observable GWs over a vast range of frequencies.
This prediction, however, relies on the assumption of a strong axion coupling,
$M_{\rm Pl}/\Lambda \sim \mathcal{O}\left(100\right)$, such that the energy
stored in the gauge field begins
to dominate the total energy budget towards the end of inflation.
As long as the backreaction from the gauge field on the inflationary dynamics
remains at a perturbative level at all times, a significantly weaker GW signal is expected.


\subsection{Baryogenesis from decaying (hyper)magnetic helicity}
\label{subsec:intro_BAU}


The prospect of a sizable GW signal from pseudoscalar inflation entails
the question as to what other observable signatures one might hope for.
Thanks to the rich phenomenology of this inflationary scenario, it should be possible
to correlate the strength of the expected GW signal to other observables.
In particular, one would like to know in which case one should either expect a strong
or only a rather weak signal in GWs.
In this context, an interesting feature of pseudoscalar inflation supplemented by
a coupling to gauge fields is the production of primordial gauge fields towards the end of
inflation~\cite{Turner:1987bw,Garretson:1992vt,Anber:2006xt}.
In fact, if the gauge symmetry $G_{\rm gauge}$ is identified with the standard model
hypercharge gauge group, $U(1)_Y$, the primordial hypermagnetic fields generated during
inflation might act as seeds
for the ubiquitous, intergalactic magnetic fields that permeate our Universe
today~\cite{Widrow:2011hs,Durrer:2013pga}.
Interestingly enough, deficits of secondary cascade photons from TeV blazars
have recently been identified, which can be explained by intergalactic magnetic
fields~\cite{Neronov:1900zz,Tavecchio:2010mk,Dolag:2010ni,Essey:2010nd,
Taylor:2011bn,Takahashi:2013lba,Finke:2015ona}.
Pseudoscalar inflation coupled to the standard model hypercharge sector,
therefore, offers an exciting opportunity for
\textit{primordial magnetogenesis}~\cite{Kandus:2010nw},
which can in principle be tested by more detailed observations of
intergalactic magnetic fields.


Moreover, the primordial (hyper)magnetic fields generated during pseudoscalar inflation
allow to generate a primordial baryon asymmetry around the time of
EWSB~\cite{Giovannini:1997gp,Giovannini:1997eg,Bamba:2006km}.
The key ingredient in this scenario of baryogenesis is the chiral triangle anomaly
in the standard model, which relates changes in the global baryon number $B$
as well as in the global lepton number $L$ to changes in the Chern-Simons
numbers in the electroweak sector,%
\footnote{Both baryon and lepton number also exhibit a gravitational anomaly,
which can likewise be used to construct scenarios of
baryogenesis~\cite{Alexander:2004us,Maleknejad:2016dci}.
In our analysis, the gravitational anomaly will, however, play no role.\smallskip}
\begin{align}
\Delta B = \Delta L = N_g \left(\Delta N_{\rm CS}^W - \Delta N_{\rm CS}^Y \right) \,,
\quad \Delta N_{\rm CS}^Y = \frac{g_Y^2}{16\pi^2}\,\Delta \mathcal{H} \,.
\label{eq:anomaly}
\end{align}
Here, $N_g = 3$ denotes the number of fermion generations in the standard model,
while $N_{\rm CS}^W$ and $N_{\rm CS}^Y$ stand for the Chern-Simons numbers associated
with the $SU(2)_W$ and $U(1)_Y$ gauge fields, respectively.%
\footnote{Of course, only $N_{\rm CS}^W$ represents a Chern-Simons number in the actual
sense, for only the weak isospin gauge sector with non-Abelian gauge group $SU(2)_W$
possesses a topologically nontrivial vacuum structure.
In the hypercharge gauge sector, the Chern-Simons number $N_{\rm CS}^Y$ is, by contrast,
understood to be related to the hypermagnetic helicity $\mathcal{H}$, which accounts
for topologically nontrivial configurations (knots) of the hypermagnetic gauge field.}
Eq.~\eqref{eq:anomaly} illustrates the well-known fact
that $SU(2)_W$ instanton and sphaleron transitions, which correspond to jumps
in the non-Abelian Chern-Simons number $N_{\rm CS}^W$, violate both $B$ and $L$.
But at the same time, Eq.~\eqref{eq:anomaly} also indicates that both
$B$ and $L$ can be generated (or destroyed) by changes in the hypermagnetic
helicity $\mathcal{H}$.
And in fact, in the presence of a maximally helical hypermagnetic field
generated during pseudoscalar inflation, this is exactly what happens
at temperatures around the electroweak scale:
The hypermagnetic field is converted into the electromagnetic (EM) field and, as a consequence,
the helicity carried by the hypermagnetic field is transferred to the one carried by
the EM field.
This corresponds to the decay of the net hypermagnetic
helicity $\mathcal{H}$, which, in turn, generates a nonzero
baryon number according to the relation in Eq.~\eqref{eq:anomaly}.
This mechanism of \textit{baryogenesis via primordial (hyper)magnetic fields}
has recently received quite some attention in the
literature~\cite{Anber:2015yca,Cado:2016kdp,Kamada:2016cnb,Kamada:2016eeb,Fujita:2016igl}
(see also \cite{Zadeh:2016nfk,Semikoz:2016lqv}).


In the following, we will adopt the results of~\cite{Kamada:2016cnb},
which represents the most comprehensive study of this scenario of
baryogenesis at the electroweak scale so far.
The authors of~\cite{Kamada:2016cnb} use recent results
from magnetohydrodynamic (MHD) simulations~\cite{Durrer:2013pga,Brandenburg:2004jv}
to model the evolution of the magnetic field.
In particular, they account for the inverse cascade behavior of the magnetic field
below a certain critical temperature, which is characterized by the transfer of power
from small scales to large scales~\cite{Frisch:1975aa,Pouquet:1976zz,Kahniashvili:2012uj}.
Moreover, they include all of the standard model Yukawa
interactions as well as the chiral magnetic effect~\cite{Vilenkin:1980fu,Anber:2006xt}.
This is essential to correctly assess the efficiency of $SU(2)_W$ sphaleron processes
in washing out the previously generated baryon number.
Finally, the authors of~\cite{Kamada:2016cnb} model the gradual conversion of
the hypermagnetic field into an EM field during EWSB, i.e., during the electroweak crossover,
$\bm{B}_Y \rightarrow \bm{B}_{\rm EM}$, in terms of a temperature-dependent weak
mixing angle $\theta_W\left(T\right)$.
In this respect, the analysis in~\cite{Kamada:2016cnb} differs drastically
from related works, which simply assume that both the generation of baryon number
as well as the $SU(2)_W$ sphalerons shut off simultaneously at temperatures around
the electroweak scale.
As shown in~\cite{Kamada:2016cnb}, this assumption turns out to be an oversimplification,
which basically corresponds to treating the electroweak crossover as a first-order phase transition.
In actual fact, the conversion of the hypermagnetic field into the EM field is
accompanied by a strong variation in the hypermagnetic helicity and, thus,
responsible for an enhanced generation of baryon number.
Likewise, one must take into account that also the emerging EM field still participates
in redistributing the total baryon number, as it communicates $B$ violation
in the left-handed fermions to the right-handed fermions.
Taken all together, the authors of~\cite{Kamada:2016cnb} find that successful
baryogenesis is feasible, as long as the present-day magnetic
field exhibits a certain physical strength, $B_p^0$, as well as a certain
physical correlation length, $\lambda_p^0$,
\begin{align}
B_p^0 \sim 10^{-17}\cdots10^{-16} \,\textrm{G} \,, \quad
\lambda_p^0 \sim  10^{-3}\cdots10^{-2} \,\textrm{pc} \,,
\label{eq:benchmark}
\end{align}
and a positive maximal helicity.
Note that these values satisfy the relation one expects for
magnetic fields that undergo the direct/inverse cascade process,
$B_p^0 = 10^{-14}\,\textrm{G}\left(\lambda_p^0/0.3\,\textrm{pc}\right)$~\cite{Banerjee:2004df}.%
\footnote{In this paper, we are going to work in \textit{natural Lorentz-Heaviside units},
in which $\hbar = c_0 = \epsilon_0 = 1$.
These are the typical units of particle physics, where the electrical
charge $e$ is supposed to be related to the fine structure constant $\alpha$
as $e = \sqrt{4\pi\alpha}$.
This means that $1\,\textrm{G} = 6.91 \times 10^{-20}\,\textrm{GeV}^2
\left(4\pi\epsilon_0\right)^{-1/2} \left(\hbar c_0\right)^{-3/2} =
1.95 \times 10^{-20}\,\textrm{GeV}^2$.
In \textit{natural Gaussian CGS units}, one has by contrast
$\hbar = c_0 = 4\pi\epsilon_0 = 1$, such that $e = \sqrt{\alpha}$
and $1\,\textrm{G} = 6.91 \times 10^{-20}\,\textrm{GeV}^2$.
To convert from our units to CGS units, one simply has
to replace $1\,\textrm{G} \rightarrow \left(4\pi\right)^{-1/2}\,\textrm{G}$.
Meanwhile, the conversion from parsec to inverse GeV is unambiguous and identical
in both unit systems, $1\,\textrm{pc} = 1.56\times 10^{32}\,\textrm{GeV}^{-1}$.
\label{fn:units}}
At the same time, they, however, come with an uncertainty of at least one
order of magnitude because of the current theoretical uncertainties in modeling
the exact evolution of the electroweak crossover.
In the following, we will use the numbers in Eq.~\eqref{eq:benchmark} as a benchmark,
keeping in mind that they merely convey an idea of the correct orders of magnitude.
Besides that, our final results can be readily carried over to other values of $B_p^0$.


\subsection{Correlation between and successful baryogenesis}
\label{subsec:synopsis}


As outlined in Sec.~\ref{subsec:intro_GWs}, pseudoscalar inflation anomalously coupled
to the gauge fields of some gauge group $G_{\rm gauge}$ results
in the enhanced production of primordial GWs.
Here, the identification of $G_{\rm gauge}$ with some non-Abelian group
results in the scenario of chromo-natural inflation~\cite{Adshead:2012kp,Adshead:2013nka}.
The description of an inflaton coupling to non-Abelian fields
is, however, slightly more challenging; and hence we shall focus on
the Abelian case in this work, for simplicity.
Furthermore, among all conceivable Abelian gauge groups that the inflaton
could couple to, the standard model hypercharge, $U(1)_Y$, certainly
plays a preeminent role.
With $U(1)_Y$ being the only Abelian gauge group in the standard model, an inflaton
coupling to the hypercharge sector may be regarded as a
\textit{most minimal} departure from the standard model.
A coupling to any other gauge symmetry, such as, e.g., $U(1)_{B-L}$, would by contrast
require the introduction of new gauge degrees of freedom (DOFs).
Moreover, coupling pseudoscalar inflation to the hypercharge sector also
offers an intriguing possibility for primordial magnetogenesis, which can be tested
by the observations of the present intergalactic magnetic fields, as well as
for baryogenesis from the decay of (hyper)magnetic helicity;
see the discussion in Sec.~\ref{subsec:intro_BAU}.
For these reasons, we deem the identification $G_{\rm gauge} \rightarrow U(1)_Y$
the most interesting choice.
In contrast to any hidden gauge symmetries beyond the standard model,
an inflaton coupling to $U(1)_Y$ is slightly less speculative and,
at the same time, more predictive in terms of observable consequences.


In this paper, we are, therefore, going to focus on general models of pseudoscalar
inflation supplemented by a Chern-Simons-type interaction between the inflaton and
the hypermagnetic gauge field.
In particular, we are going to address the following two questions:


(1) Under what conditions does pseudoscalar inflation result in a
(hyper)magnetic field of just the right magnitude, such that primordial magnetogenesis
at the end of inflation sets the stage for successful baryogenesis at the electroweak scale?
That is, how does one need to choose the parameters of pseudoscalar inflation
in order to satisfy the two conditions in Eq.~\eqref{eq:benchmark}?
In this part of our analysis, we are basically going to update previous studies
of \textit{baryogenesis from pseudoscalar inflation}~\cite{Anber:2015yca,Cado:2016kdp}
(see also \cite{Adshead:2016iae}).
By employing the results presented in~\cite{Kamada:2016cnb}, we make sure to
include several important effects that had been neglected up to this point
(such as, e.g., the inverse cascade regime, the chiral magnetic effect, and
the role of the standard model Yukawa interactions).
In doing so, we will work in the approximation of instant reheating, for simplicity.
In principle, both magnetic fields and gravitational waves are also produced
during the stage of reheating~\cite{Adshead:2015pva,Fujita:2015iga,Adshead:2016iae}.
The correct description of this phase, however, requires a dedicated
numerical simulation that includes both nonperturbative particle production
and MHD.
In particular, one should take into account the backreaction on the gauge field
production from the hypercharged particles in the emerging plasma.
Such a study is not yet available, which is why we will ignore the details
of the reheating phase altogether.
On the one hand, the approximation of instant reheating introduces
some (perhaps very large) uncertainties into our analysis.%
\footnote{The lattice simulation in~\cite{Adshead:2016iae}, e.g., indicates
a large enhancement of hypermagnetic fields at the stage of reheating.
On the contrary, the authors of \cite{Fujita:2015iga} point out the necessity of a
relatively low reheating temperature in order to avoid high electric conductivity,
which would otherwise prevent hypermagnetic helicity from developing during reheating.
However, a low reheating temperature automatically
comes with a large dilution of the hypermagnetic field.
From this perspective, one would therefore rather
expect a suppression than an enhancement from reheating.
In the following, we will evade the (still on-going)
debate which of these conclusions is correct and simply
neglect any contributions to the hypermagnetic field from reheating.
Instead, we will simply focus on the gauge field production
during inflation.
In this sense, our estimate is a quantitatively conservative one.}
On the other hand, it allows us to remain absolutely model-independent,
as far as the concrete dynamics of inflation and reheating are concerned.
Against this background, we hope that our analysis may motivate further studies of
reheating after pseudoscalar inflation
that account for the complicated interplay between gauge field production
and the properties of the emerging charged plasma.


(2) What are the implications of successful baryogenesis for the present-day
GW spectrum?
Assuming that primordial magnetogenesis results in magnetic fields in accord with
Eq.~\eqref{eq:benchmark}, is there still a chance to obtain GWs that could
be detected in GW experiments in the near future?


To answer these questions, we will now proceed as follows:
In Sec.~\ref{sec:magnetogenesis}, we will first review the production of hypermagnetic
fields in models of pseudoscalar inflation.
We will discuss in particular the dependence on the suppression scale $\Lambda$
as well as the backreaction on the inflationary dynamics.
In Sec.~\ref{sec:evolution}, we will then study the evolution of the primordial
hypermagnetic fields from the time of their production all the way to the present epoch.
In Sec.~\ref{sec:BAUGWs}, we will in turn study the implications for baryogenesis as well
as for the GW spectrum.
Here, our main interest will be to
establish a connection between successful baryogenesis and the expected
strength of the GW signal from inflation.
In Sec.~\ref{sec:example}, we will finally illustrate some of our main results numerically
by means of a concrete example, based on the original model of
natural inflation~\cite{Freese:1990rb,Adams:1992bn}.
Sec.~\ref{sec:conclusions} contains our conclusions as well as a
brief outlook on how our work could be continued.


\section{Gauge field production during inflation}
\label{sec:magnetogenesis}


We begin by reviewing the mechanism of gauge field production
in models of pseudoscalar inflation~\cite{Turner:1987bw,Garretson:1992vt,Anber:2006xt}.
This will also serve the purpose to establish our notation and conventions.


\subsection{Equations of motion for the inflaton and gauge fields}
\label{subsec:EOMs}


For an arbitrary model of pseudoscalar inflation coupled to the standard model
hypercharge sector via an effective Chern-Simons term, the relevant Lagrangian
takes the following form,
\begin{align}
\mathcal{L} \supset -\frac{1}{2}\partial_\mu a\, \partial^\mu a
- \frac{1}{4} F_{\mu\nu} F^{\mu\nu} - V\left(a\right)
- \frac{a}{4\Lambda} F_{\mu\nu}\tilde{F}^{\mu\nu} \,.
\label{eq:Lagrangian}
\end{align}
Here, the field $a$ denotes the axion-like pseudoscalar inflaton;
$F_{\mu\nu} = \partial_\mu A_\nu - \partial_\nu A_\mu$ is the
field strength tensor belonging to the hypercharge gauge field $A_\mu$;
and $\tilde{F}^{\mu\nu}$ is the dual field strength tensor,
$\tilde{F}^{\mu\nu} = \frac{1}{2}\left(-g\right)^{-1/2} \epsilon^{\mu\nu\sigma\tau}F_{\sigma\tau}$.
For the time being, we remain as model-independent as possible and
do not specify the concrete form of the inflaton potential $V\left(a\right)$.
Only in Sec.~\ref{sec:example}, we will become more explicit and identify
$V\left(a\right)$ with the scalar potential of particular models of inflation.
The last term in Eq.~\eqref{eq:Lagrangian} represents the anomalous
Chern-Simons interaction between the inflaton and the hypercharge gauge field.
The parameter $\Lambda$ denotes an effective suppression scale, the magnitude of which
is related to the energy scale at which the anomalous coupling is generated.
In the following, we will treat it as a free parameter.
The combination $a/\Lambda$, i.e., the prefactor of the topological term
$\frac{1}{4} F_{\mu\nu}\tilde{F}^{\mu\nu}$,
may be regarded as an effective, field-dependent vacuum angle in the
hypercharge sector, $\theta = a/\Lambda$.
If we replaced $a$ by a constant, $\theta$ would become unphysical
and could be transformed away by a fermion rotation.
However, with $a$ being a dynamical field, the vacuum
angle $\theta$ is physically meaningful; see also~\cite{Perez:2014fja}.


Given the Lagrangian in Eq.~\eqref{eq:Lagrangian}, one obtains for
the homogeneous Friedmann equation,
\begin{align}
H^2 = \bigg(\frac{\dot{R}}{R}\bigg)^2 = \frac{\rho}{3M_{\rm Pl}^2} \,, \quad
\rho = \frac{1}{2}\dot{a}^2 + V\left(a\right) +
\frac{1}{2}\left<\bm{E}^2\right> + \frac{1}{2}\left<\bm{B}^2\right> \,.
\label{eq:Friedmann}
\end{align}
Here, $H$ is the Hubble rate; $R$ denotes the scale factor in the
Friedmann-Lema\^itre-Robertson-Walker metric,
$ds^2 = -dt^2 + R^2\left(t\right)d\bm{x}^2 = - R^2\left(t\right)\left(d\tau^2-d\bm{x}^2\right)$;
$\rho$ represents the total energy density;
and $M_{\rm Pl} = \left(8\pi G\right)^{-1/2} = 2.44 \times 10^{18}\,\textrm{GeV}$
is the reduced Planck mass.
$\dot{R}$ stands for the derivative of the scale factor w.r.t.\ physical time $t$.
Below, we will also encounter derivatives w.r.t.\ conformal time $\tau$,
which will be denoted by a prime.
The total energy density $\rho$ can be obtained from the stress-energy tensor.
In addition to the usual contributions from the inflaton field, it now also
receives contributions from the hyperelectric and hypermagnetic fields $\bm{E}$ and $\bm{B}$.
We are going to work in radiation gauge, which combines the gauge fixing
conditions of Coulomb (or transverse) gauge, $\bm{\nabla}\cdot \bm{A} = 0$, and
Weyl (or temporal) gauge, $A_0 = 0$.
The fields $\bm{E}$ and $\bm{B}$ are then related to the
components of the hypercharge vector field $A_\mu$ as follows,
\begin{align}
A_\mu = \left(A_0,\bm{A}\right) \,, \quad
\bm{E} = - \frac{1}{R^2}\,\partial_\tau \bm{A} = - \frac{1}{R^2}\,\bm{A}' \,, \quad
\bm{B} =   \frac{1}{R^2}\,\bm{\nabla}\times \bm{A} \,.
\end{align}
$\bm{E}$ and $\bm{B}$ are understood to represent \textit{physical} field
strengths, whereas $A_\mu$ is a \textit{comoving} quantity that needs to be
determined in dependence of the comoving coordinates $x^\mu = \left(\tau,\bm{x}\right)$.
The angle brackets in Eq.~\eqref{eq:Friedmann} denote the expectation
values of $\bm{E}^2$ and $\bm{B}^2$, respectively.
During inflation, these expectation values correspond to \textit{quantum mechanical}
vacuum expectation values.
In order to determine the \textit{classical} field strengths after inflation,
we identify these quantum expectation values with the ensemble averages of the classical
fields just at the end of inflation,
\begin{align}
\left<\cdot\right>_{\textrm{quantum vacuum}}
\quad\overset{\textrm{end of inflation}}{\longrightarrow}\quad
\left<\cdot\right>_{\textrm{classical ensemble}} \,.
\end{align}


Similarly as the Friedmann equation, the equation of motion for the homogeneous inflaton field
also turns out to receive corrections in presence of the anomalous axion-gauge-field coupling,
\begin{align}
\ddot{a} + 3 H \dot{a} + \frac{dV}{da} = \frac{1}{\Lambda} \left<\bm{E}\bm{B}\right> \,.
\label{eq:KleinGordon}
\end{align}
Here, the new source term on the right-hand side may be regarded as  an additional friction
term (next to the usual Hubble friction term, $3 H \dot{a}$).
In the case of strong gauge field production, the source term eventually
dominates over the Hubble friction term, which alters the inflationary dynamics
towards the end of inflation~\cite{Anber:2009ua,Barnaby:2011qe,Barnaby:2011vw}
(see also \cite{Domcke:2016bkh,Garcia-Bellido:2016dkw}).
As we will see later on, this regime will be less relevant for our purposes, i.e.,
as long as we require successful baryogenesis.


The dynamics of the vector field are governed by the following wave equation,
\begin{align}
\Box \bm{A} =
-\bm{A}'' + \bm{\nabla}^2 \bm{A} = - \frac{a'}{\Lambda}\, \bm{\nabla} \times \bm{A} \,,
\label{eq:AEOM}
\end{align}
where the axion-gauge-field coupling induces again a source term on the right-hand side.
To find the solution of this equation, it is convenient to perform a Fourier transform
and work in momentum space.
Upon quantization of the individual Fourier modes, $\bm{A}$ may
be written as
\begin{align}
\bm{A}\left(\tau,\bm{x}\right) = \sum_{\lambda=\pm}
\int \frac{d^3\bm{k}}{\left(2\pi\right)^{3/2}}\left[
A_\lambda\left(\tau,\bm{k}\right)\bm{\epsilon}_\lambda\left(\bm{k}\right)
\hat{a}_\lambda\left(\bm{k}\right) e^{i\bm{k}\bm{x}} +\textrm{h.c.}\right] \,.
\label{eq:Fourier}
\end{align}
Here, $\lambda = \pm$ labels the two possible helicity states;
$A_\pm$ denote the corresponding mode functions; $\bm{\epsilon}_\pm$ are
the two polarization vectors; and $\hat{a}_\pm$ stand for the corresponding
annihilation operators, which annihilate states $\left|\bm{k},\lambda\right>$
with 3-momentum $\bm{k}$ and polarization $\lambda$.
The vectors $\bm{\epsilon}_\pm$ for given momentum $\bm{k}$
form an orthonormal basis in the complex vector space perpendicular to $\bm{k}$,
\begin{align}
\bm{\epsilon}_\lambda\left(\bm{k}\right) \cdot \bm{\epsilon}_{\lambda'}^*\left(\bm{k}\right)
= \delta_{\lambda\lambda'} \,, \quad
\bm{\epsilon}_\lambda\left(\bm{k}\right) \cdot \bm{k} = 0 \,, \quad
i\bm{k}\times\bm{\epsilon}_\lambda\left(\bm{k}\right) =
\lambda\, k\, \bm{\epsilon}_\lambda\left(\bm{k}\right) \,,
\label{eq:epsilonk}
\end{align}
where $k = \left|\bm{k}\right|$.
Meanwhile, the annihilation and creation operators,
$\hat{a}_\lambda\left(\bm{k}\right)$ and $\hat{a}_\lambda^\dagger\left(\bm{k}\right)$,
satisfy the usual canonical commutation relations,
$\big[\hat{a}_\lambda\left(\bm{k}\right),\hat{a}_{\lambda'}^\dagger\left(\bm{k}'\right)\big]
= \delta_{\lambda\lambda'}\,\delta^{(3)}\left(\bm{k}-\bm{k}'\right)$.
Inserting the Fourier expansion in Eq.~\eqref{eq:Fourier} into the
equation of motion in Eq.~\eqref{eq:AEOM} and using the relations
in Eq.~\eqref{eq:epsilonk}, one then obtains the following mode equations
in momentum space,
\begin{align}
\left[\frac{\partial^2}{\partial\tau^2} + k^2
\left(1 - \frac{x_\lambda\left(\xi\right)}{x\left(\tau,k\right)}\right)\right]
A_\lambda\left(\tau,\bm{k}\right) = 0 \,, \quad
x\left(\tau,k\right) = -k\tau \,, \quad x_\lambda\left(\xi\right) = 2\lambda\,\xi \,,
\label{eq:modeequation}
\end{align}
where we have defined the \textit{instability parameter} $\xi$ as follows,
\begin{align}
\xi = \frac{1}{2H} \frac{\dot{a}}{\Lambda} \,.
\label{eq:xi}
\end{align}
The mode equations are isotropic in momentum space, which is why we will
label the mode functions only by their absolute momenta
from now on, $A_\lambda\left(\tau,\bm{k}\right) \rightarrow A_\lambda^k\left(\tau\right)$.
The parameter $x$ in Eq.~\eqref{eq:modeequation} quantifies whether,
at a certain conformal time $\tau$, a given mode with wavenumber $k$ has a spatial extent
(i.e., physical wavelength $\lambda_p = 2\pi R/k$)
larger or smaller than the Hubble radius, $H^{-1}$.
To see this, one simply has to recall that during inflation,
i.e., in quasi-de Sitter space, $\tau$ is approximately given as
$\tau \simeq - 1 /\left(RH\right)$.
This readily implies $x \simeq 2\pi H^{-1} / \lambda_p$.
The magnitude of $x$ needs to be compared with $x_\lambda$,
which is defined in terms of the instability parameter $\xi$.
The parameter $x_\lambda = \lambda\, \dot{\theta}/H$ in Eq.~\eqref{eq:modeequation} hence
measures the rate of variation of the effective vacuum angle $\theta = a/\Lambda$
in relation to the Hubble rate $H$.
With the above definitions, one also finds that
$x_\lambda/x = \lambda\, k_{\rm crit} / k$, where $k_{\rm crit} = R\, \dot{\theta}$
is a certain critical (comoving) momentum scale.
From the perspective of gauge field production, the quantities
$x_\lambda$, $\xi$, and $k_{\rm crit}$ vary only slowly with time.
This is a direct consequence of the slow-roll motion of the field $a$ during inflation.
When solving the mode equations in Eq.~\eqref{eq:modeequation},
we will, therefore, treat $x_\lambda$ at any given moment in time as a constant.
This will provide us with solutions for the vector-field modes
that are respectively valid during certain periods of inflation,
when $x_\lambda$ takes particular, approximately constant values.
Other than that, we will make no further approximations when solving
Eq.~\eqref{eq:modeequation}.


From Eq.~\eqref{eq:modeequation}, it is evident that, for $x<\left|x_\lambda \right|$,
the helicity modes corresponding to positive $x_\lambda$ become tachyonically unstable.
A positive baryon asymmetry requires a positive (hyper)magnetic
helicity~\cite{Fujita:2016igl,Kamada:2016eeb,Kamada:2016cnb}.
In the following, we will therefore consider the case where $\dot{a}>0$, such
that $x_+ > 0$ and $x_- < 0$.
In this case, the positive-helicity modes  $A_+^k$ will be tachyonically unstable at $x<x_+$.%
\footnote{Conversely, in the case of negative inflaton velocity, $\dot{a}<0$,
we would have to deal with $x_- > 0$ and $x_+ < 0$.
This would result in a negative helicity and, consequently, in a negative baryon asymmetry.
For the inflationary dynamics, the sign of the induced helicity does not matter.
Moreover, as long as the inflaton potential is invariant under parity, $a \leftrightarrow -a$,
the sign of the inflaton velocity does not affect the inflationary dynamics as well.}
Once $x$ has dropped down to values smaller than $x_+$,
the modes $A_+^k$ begin to exponentially grow.
The negative-helicity modes $A_-^k$ experience, by contrast, only a shift
in their dispersion relation towards effectively \textit{larger} momenta,
$k^2 \rightarrow k^2 \big(1+k_{\rm crit}/k\big)$.
They, thus, always stay at the quantum level.
For constant $\xi$, the exact solutions for $A_\pm^k$ are given in terms of Whittaker $W$
functions (which are related to confluent hypergeometric functions)~\cite{Abramowitz:1972aa}.
This is because Eq.~\eqref{eq:modeequation} can be brought
into a particular form of Whittaker's equation,
\begin{align}
\left(\frac{d^2}{dz^2} - \frac{1}{4}
+ \frac{\kappa_\lambda}{z}\right)
A_\lambda^k\left(z\right) = 0 \,, \quad
z =  -2ix = 2ik\tau \,, \quad
\kappa_\lambda = \frac{x_\lambda}{2 i} = -i\lambda\,\xi \,.
\label{eq:Whittaker}
\end{align}
We require that the modes $A_\pm^k$ reduce to the usual
Bunch-Davis solution in the asymptotic past,
\begin{align}
\lim_{-k\tau\rightarrow \infty} A_\lambda^k\left(\tau\right) = \frac{e^{-ik\tau}}{\sqrt{2k}} \,.
\end{align}
With this boundary condition, the Whittaker equation in Eq.~\eqref{eq:Whittaker}
has the following solution,
\begin{align}
A_\lambda^k\left(\tau\right) =
\frac{e^{\lambda\pi\xi/2}}{\sqrt{2k}}\,
W_{-i\lambda\,\xi,1/2}\left(2ik\tau\right) \,,
\label{eq:Asolution}
\end{align}
where $W_{-i\lambda\,\xi,1/2}$ is the Whittaker function $W_{\kappa,\mu}$
with indices $\kappa=\kappa_\lambda = -i\lambda\,\xi$ and $\mu = 1/2$.
This function grows exponentially as a function of $\tau$ for $\lambda = +$
and remains oscillatory for $\lambda = -$.


\subsection{Backreaction on the inflationary dynamics}
\label{subsec:backreaction}


In the previous section, we have seen how the axion-induced source term on the
right-hand side of Eq.~\eqref{eq:AEOM} manages to excite vector-field modes
with positive helicity; see Eq.~\eqref{eq:Asolution}.
We shall now examine the consequence of this nonperturbative
gauge field production for the inflationary dynamics.
In the presence of a macroscopic gauge field configuration, the Friedmann and
Klein-Gordon equations in Eqs.~\eqref{eq:Friedmann} and \eqref{eq:KleinGordon}
need to be supplemented by the following expressions,
\begin{align}
\label{eq:rhoEB}
\rho_{EE} & = \frac{1}{2}\left<\bm{E}^2\right>
= \frac{1}{2R^4} \int \frac{d^3\bm{k}}{\left(2\pi\right)^3}
\left|\frac{\partial}{\partial \tau} A_+^k\right|^2
\,, \\ \nonumber
\rho_{BB} & = \frac{1}{2}\left<\bm{B}^2\right>
= \frac{1}{2R^4} \int \frac{d^3\bm{k}}{\left(2\pi\right)^3}
\, k^2 \left|A_+^k\right|^2
\,, \\ \nonumber
\rho_{EB} & = \frac{1}{2}\left<\bm{E}\bm{B}\right> + \frac{1}{2}\left<\bm{B}\bm{E}\right>
= - \frac{1}{2R^4} \int \frac{d^3\bm{k}}{\left(2\pi\right)^3}
\, k\,\frac{\partial}{\partial \tau} \left|A_+^k\right|^2
\,,
\end{align}
where we neglect the vacuum contributions from the negative-helicity modes.
The quantities $\rho_{EE}$ and $\rho_{BB}$ have a direct interpretation
in the sense that they correspond to the energy densities stored in the hyperelectric
and hypermagnetic fields, respectively.
The quantity $\rho_{EB}$ is the corresponding cross term.
We note that the $\bm{E}$ and $\bm{B}$ fields do \textit{not} commute
at the quantum level, which is why $\rho_{EB}$ is defined as the symmetrized
version of $\left<\bm{E}\bm{B}\right>$.
Technically, the right-hand side of Eq.~\eqref{eq:KleinGordon} is understood
to correspond to $\rho_{EB}/\Lambda$.
In the classical limit, the commutator $\left[\bm{E},\bm{B}\right]$, however,
vanishes and $\rho_{EB}$ and $\left<\bm{E}\bm{B}\right>$ become
equivalent to each other.


The energy densities in Eq.~\eqref{eq:rhoEB}
are functions of the inflationary Hubble rate
$H$ as well as of the instability parameter $\xi$; see Eq.~\eqref{eq:xi}.
To extract the dependence on these two parameters, it turns out convenient
to rewrite the momentum integrals in Eq.~\eqref{eq:rhoEB} as follows,
\begin{align}
\label{eq:rhoI}
\rho_{EE}  = \mathcal{I}_{EE}\left(\xi\right)\,\frac{e^{2\pi\xi}}{\xi^3}\, H^4 \,, \quad
\rho_{BB}  = \mathcal{I}_{BB}\left(\xi\right)\,\frac{e^{2\pi\xi}}{\xi^5}\, H^4 \,, \quad
\rho_{EB}  = - \mathcal{I}_{EB}\left(\xi\right)\,\frac{e^{2\pi\xi}}{\xi^4}\, H^4 \,,
\end{align}
with the integral functions $\mathcal{I}_{EE}$, $\mathcal{I}_{BB}$ and $\mathcal{I}_{EB}$
being defined as
\begin{align}
\label{eq:integrals}
\mathcal{I}_{EE}\left(\xi\right) & = \frac{\xi^3}{8\pi^2}\, e^{-\pi\,\xi}
\int_{0}^{x_{\rm UV}} dx\,x^3 \left|\frac{\partial}{\partial x}
W_{\kappa_+,1/2}\left(-2ix\right)\right|^2  \,,
\\ \nonumber
\mathcal{I}_{BB}\left(\xi\right) & = \frac{\xi^5}{8\pi^2}\, e^{-\pi\,\xi}
\int_{0}^{x_{\rm UV}} dx\,x^3 \left|W_{\kappa_+,1/2}\left(-2ix\right)\right|^2 \,,
\\ \nonumber
\mathcal{I}_{EB}\left(\xi\right) & = \frac{-\xi^4}{8\pi^2}\, e^{-\pi\,\xi}
\int_{0}^{x_{\rm UV}} dx\,x^3\,\frac{\partial}{\partial x}
\left|W_{\kappa_+,1/2}\left(-2ix\right)\right|^2 \,.
\end{align}
Here, we choose a sign convention such that all three functions are positive.
The fact that $\rho_{EB}$ actually takes negative values
is accounted for by the explicit minus sign in Eq.~\eqref{eq:rhoI}.
In principle, the momentum integrals in Eq.~\eqref{eq:rhoEB} are UV-divergent,
as they receive vacuum contributions from an infinite number of high-frequency modes
(i.e., modes deep inside the Hubble horizon).
To regularize this divergence, we introduce a UV cut-off scale,
$x_{\rm UV} = k_{\rm UV}/\left(RH\right)$, which allows us to integrate
over only those modes that are excited above the vacuum level.
The natural choice for $x_{\rm UV}$ is consequently $x_{\rm UV} = x_+ = 2\xi$,
such that the momentum cut-off $k_{\rm UV}$ coincides with $k_{\rm crit}$, i.e.,
the highest wavenumber that still leads to a tachyonic instability in Eq.~\eqref{eq:modeequation}.


In view of Eq.~\eqref{eq:integrals}, it is also interesting to note
that we absorbed the \textit{explicit} time dependence of the
vector-field modes $A_\pm^k\left(\tau\right)$ in Eq.~\eqref{eq:rhoEB}
into the integration variable $x = -k\tau$.
The remaining time dependence is then
canceled by the time dependence of $R^{-4}$ in front of the integrals
in Eq.~\eqref{eq:rhoEB}.
At first glance, this renders all of the three quantities in Eq.~\eqref{eq:rhoI} constant in time.
However, there remains an \textit{implicit} time dependence encoded in the parameters $\xi$ and $H$,
which actually slowly vary during inflation.
In the following, we will determine $\rho_{EE}$, $\rho_{BB}$, $\rho_{EB}$ at any time
$t$ during inflation simply by evaluating Eq.~\eqref{eq:rhoI} for the respective values of
$\xi\left(t\right)$ and $H\left(t\right)$.
If we were to treat the time dependence of $\xi$ and $H$ more carefully, we would have to solve
Eqs.~\eqref{eq:Friedmann}, \eqref{eq:KleinGordon}, and \eqref{eq:modeequation} simultaneously.
Such an analysis is beyond the scope of this paper.

\begin{figure}[t]
\begin{center}
\includegraphics[width=0.48\textwidth]{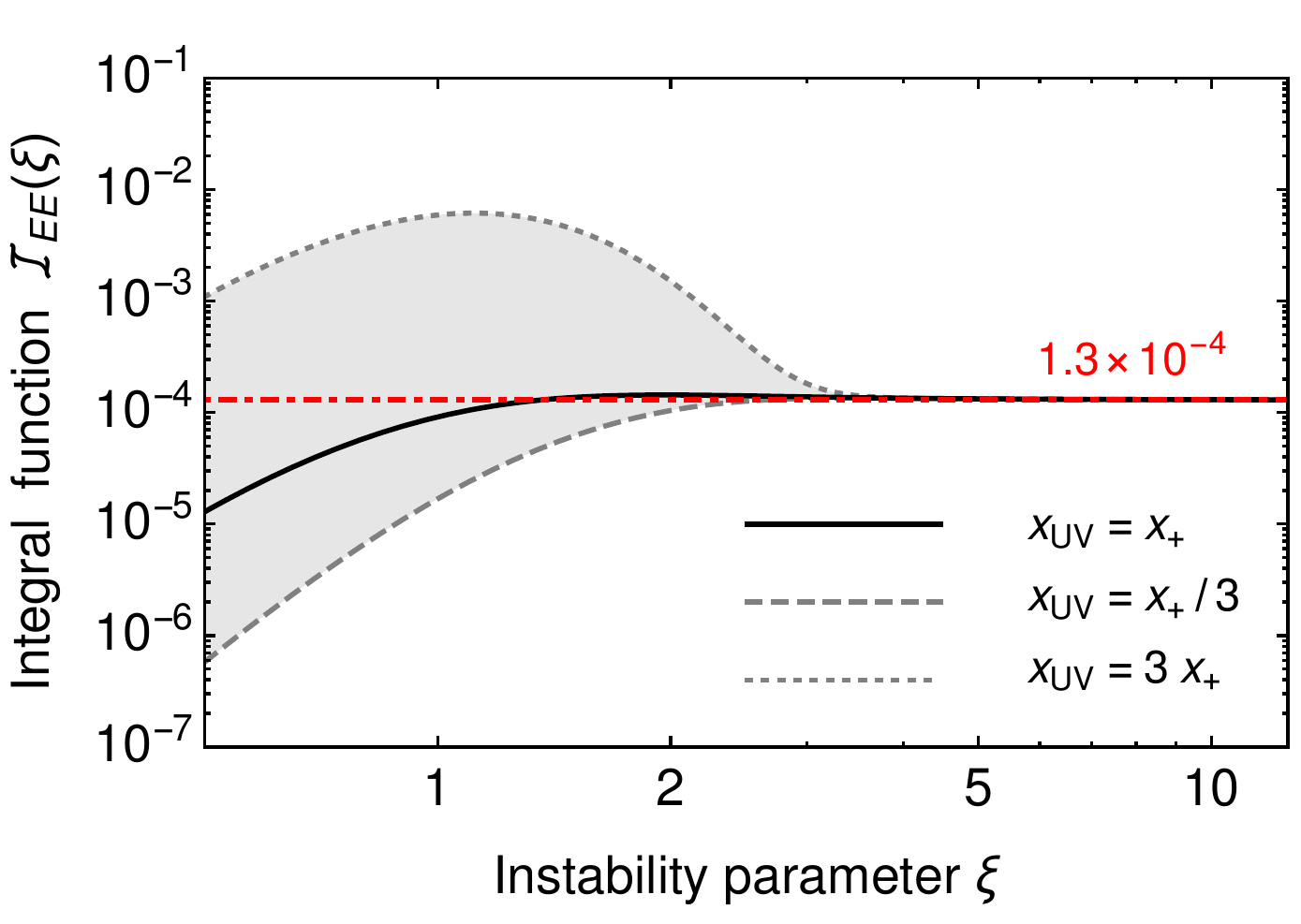}\hfill
\includegraphics[width=0.48\textwidth]{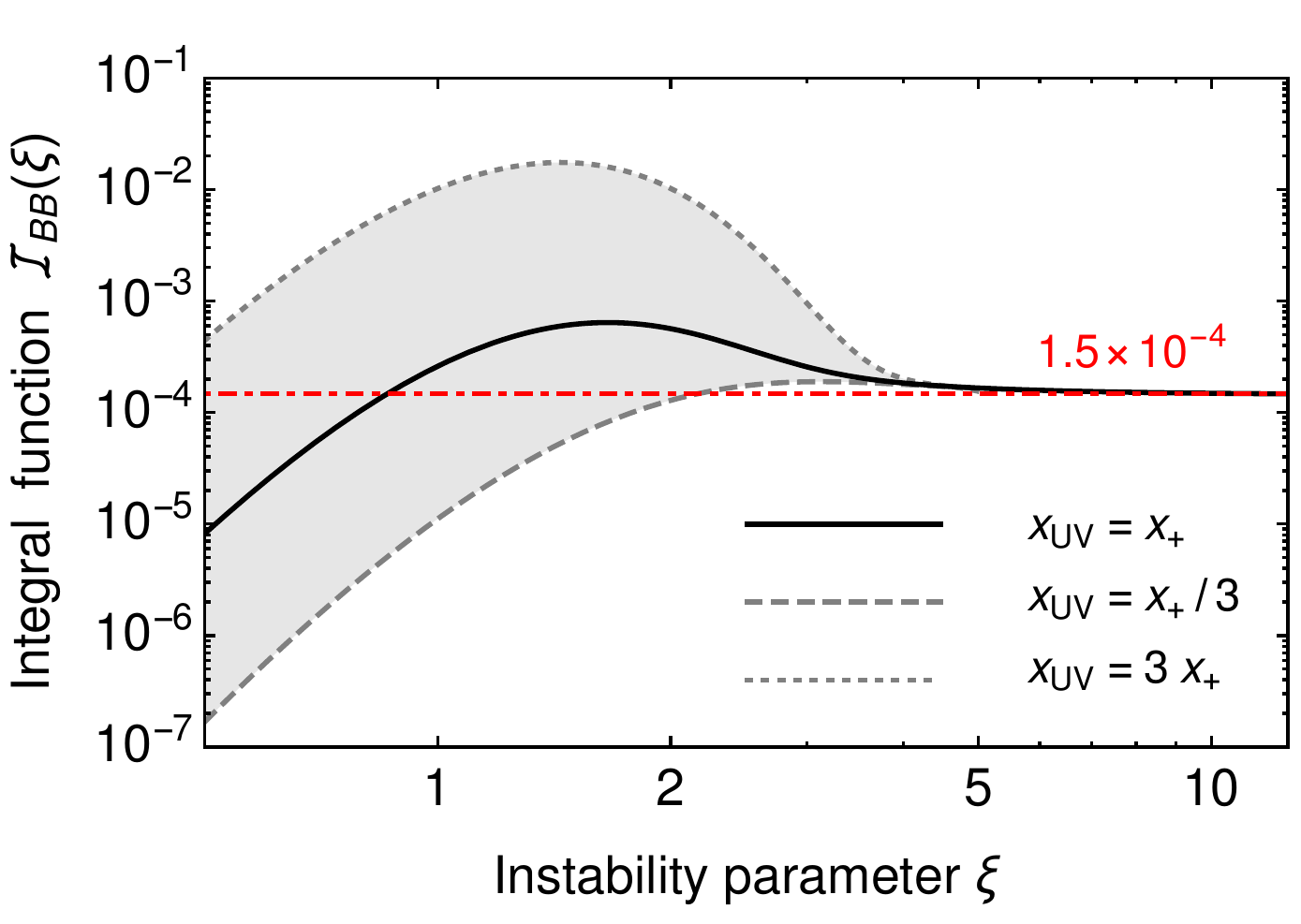}

\includegraphics[width=0.48\textwidth]{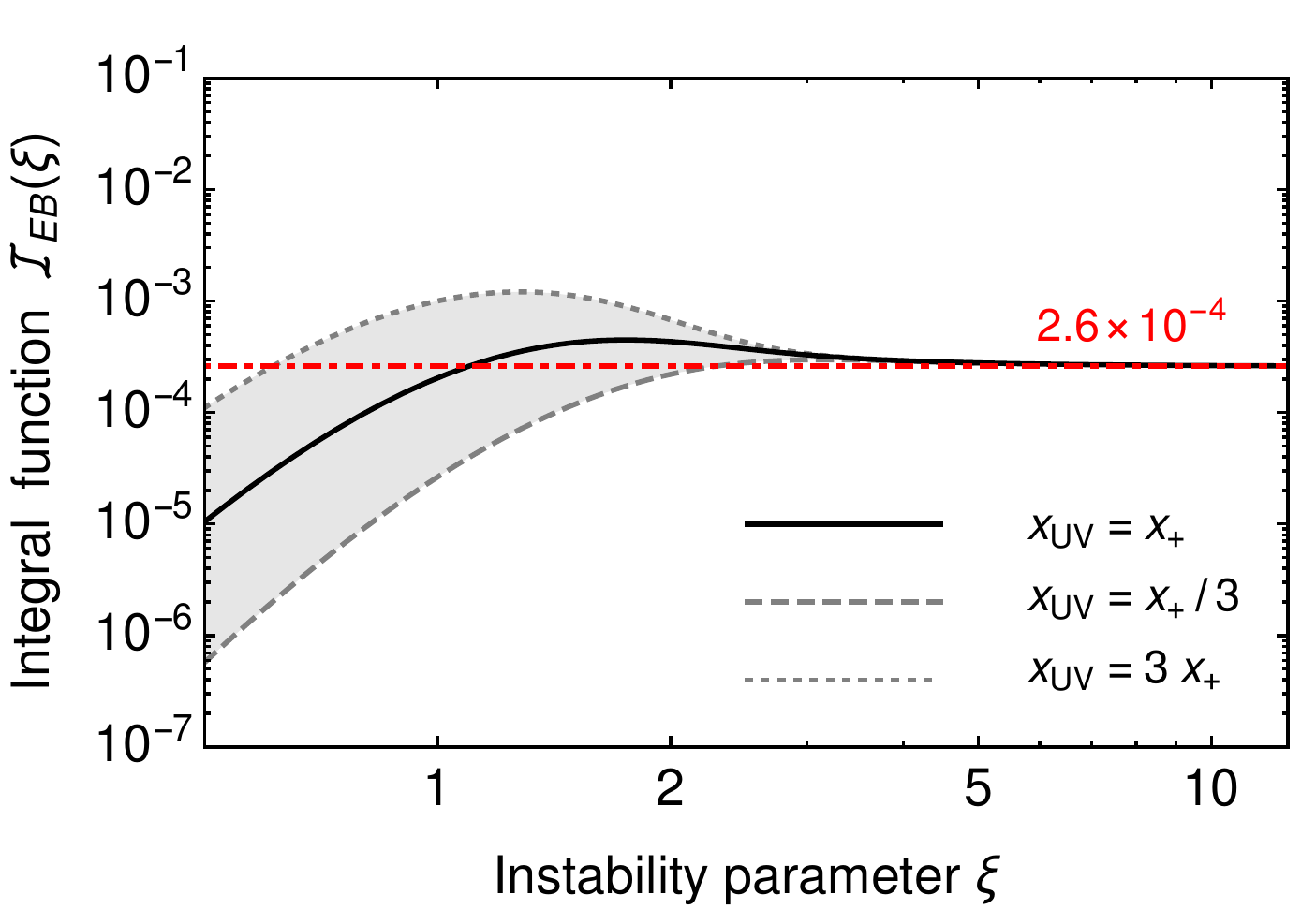}\hfill
\includegraphics[width=0.48\textwidth]{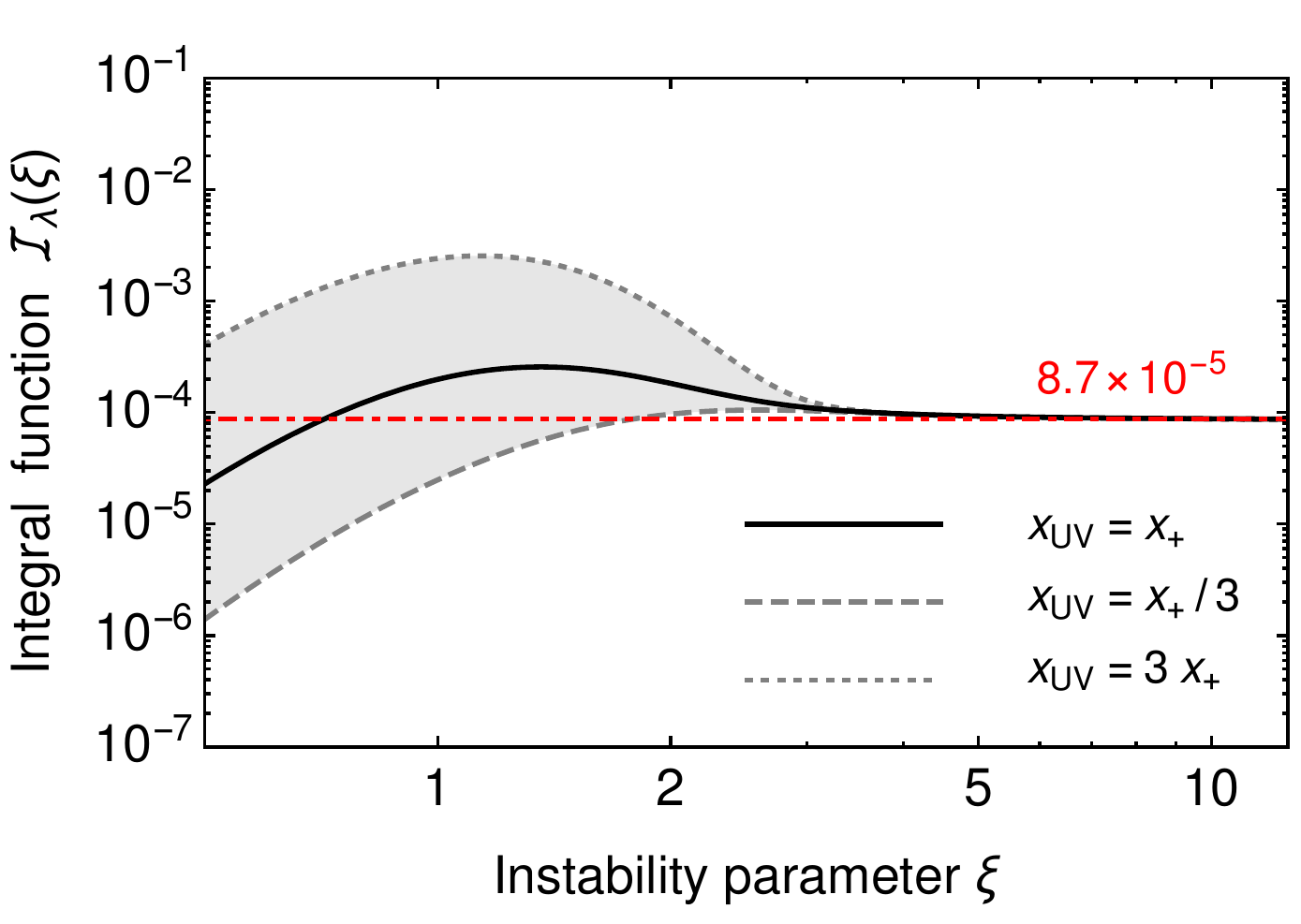}
\caption{Dependence of the integral functions
$\mathcal{I}_{EE}$ \textbf{(upper left panel)},
$\mathcal{I}_{BB}$ \textbf{(upper right panel)},
$\mathcal{I}_{EB}$ \textbf{(lower left panel)}, and
$\mathcal{I}_\lambda$ \textbf{(lower right panel)},
on the instability parameter $\xi$;
see Eqs.~\eqref{eq:xi}, \eqref{eq:integrals}, and \eqref{eq:Ilambda}.
For each function, we illustrate the effect of varying the UV cut-off scale
(parametrized in terms of the upper integration boundary $x_{\rm UV}$)
within roughly one order of magnitude.
At any given value of $\xi$, the parameter $x_+$ corresponds to $x_+ = 2\xi$;
see Eq.~\eqref{eq:modeequation}.
The red numbers and horizontal lines indicate the respective asymptotic values at $\xi \gg 1$.}
\label{fig:integrals}
\end{center}
\end{figure}

The advantage of the parametrization in Eq.~\eqref{eq:rhoI} is that all of
the three functions $\mathcal{I}_{EE}$, $\mathcal{I}_{BB}$, and $\mathcal{I}_{BE}$
asymptotically approach constant values at $\xi \gg 1$.
This is depicted in Fig.~\ref{fig:integrals}, where we also demonstrate
the sensitivity of the three integral functions to variations in the UV cut-off.
As can be seen from Fig.~\ref{fig:integrals}, all three functions become
insensitive to the exact choice for $x_{\rm UV}$ as soon as they approach
their respective asymptotic values.
For $\xi \gtrsim 4$, it is, therefore, safe to
approximate $\mathcal{I}_{EE}$, $\mathcal{I}_{BB}$, and $\mathcal{I}_{BE}$
by the constant values shown in Fig.~\ref{fig:integrals},
\begin{align}
\label{eq:rhonum}
\rho_{EE} \simeq 1.3 \times 10^{-4}\,\frac{e^{2\pi\xi}}{\xi^3}\, H^4 \,, \quad
\rho_{BB} \simeq 1.5 \times 10^{-4}\,\frac{e^{2\pi\xi}}{\xi^5}\, H^4 \,, \quad
\rho_{EB} \simeq -2.6 \times 10^{-4}\,\frac{e^{2\pi\xi}}{\xi^4}\, H^4 \,.
\end{align}
These results are consistent with the approximate solution
for the excited mode functions,
$A_+^k\left(x\right) = \left(2k\right)^{-1/2}\left(x/x_+\right)^{1/4}
\exp\left[\pi x_+/2 - 2 \sqrt{xx_+}\right]$,
which is often employed in the literature.


\begin{figure}[t]
\begin{center}
\includegraphics[width=0.48\textwidth]{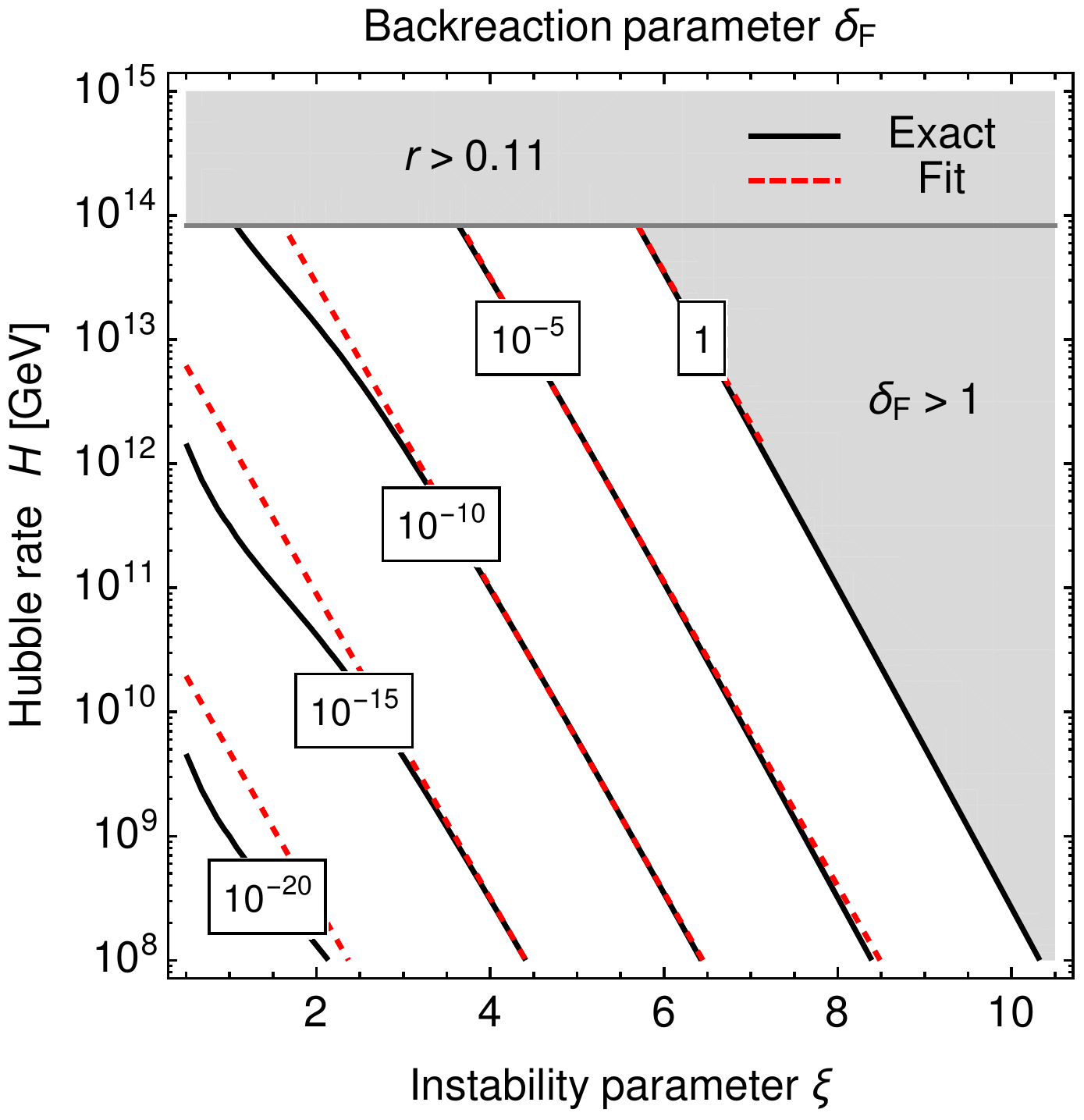}\hfill
\includegraphics[width=0.48\textwidth]{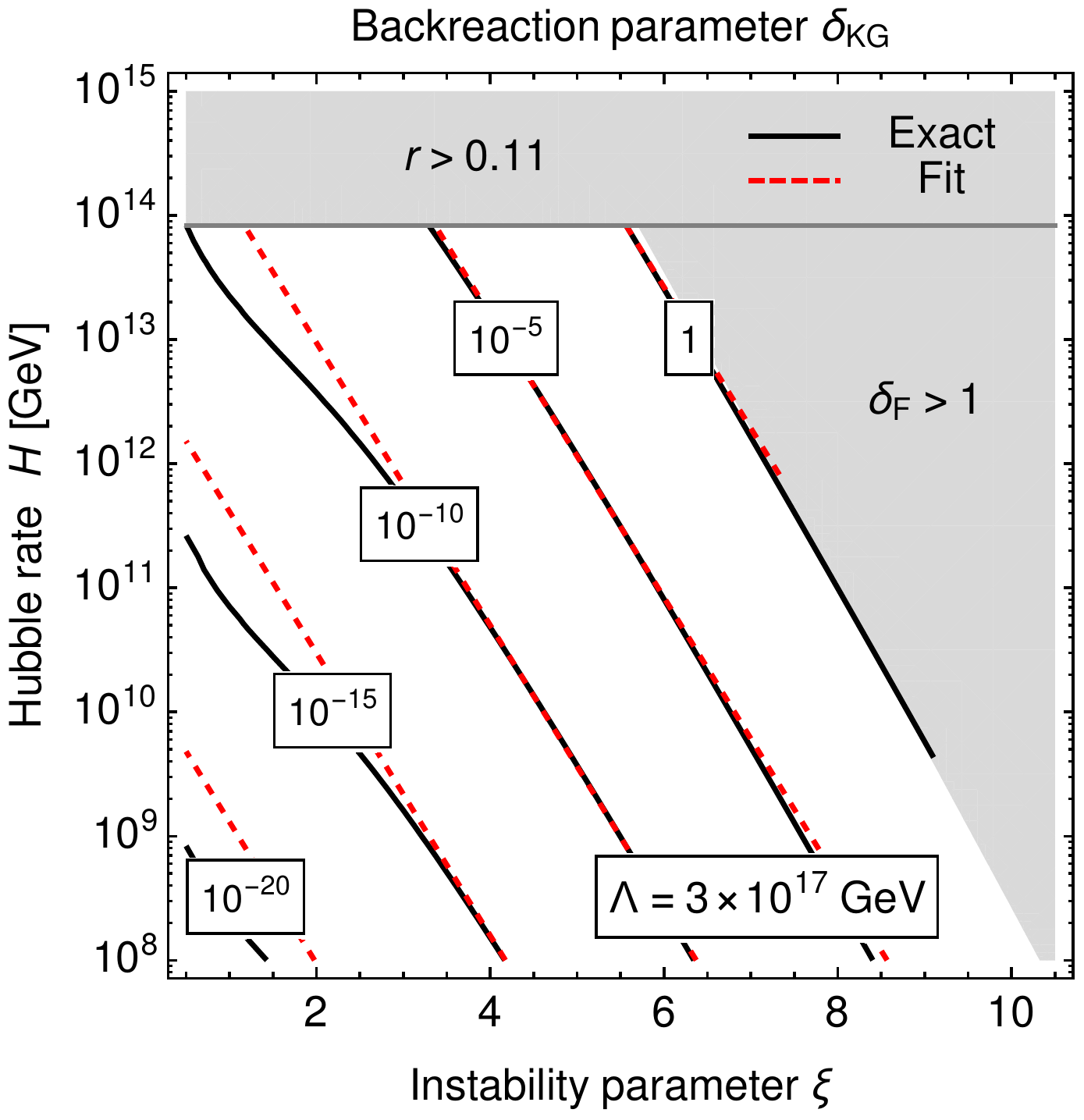}
\caption{Backreaction parameters $\delta_{\rm F}$ \textbf{(left panel)}
and $\delta_{\rm KG}$ \textbf{(right panel)}
as functions of the instability parameter $\xi$ and the Hubble rate $H$.
The parameter $\delta_{\rm F}$ quantifies the amount of backreaction
in the Friedmann equation, while the parameter $\delta_{\rm KG}$ quantifies
the amount of backreaction in the Klein-Gordon equation;
see Eq.~\eqref{eq:deltaDef}.
The black solid contours represent the exact expressions for
$\delta_{\rm F}$ and $\delta_{\rm KG}$, including the
complicated $\xi$ dependence of the integral functions
in Eq.~\eqref{eq:integrals}.
The red dashed contours
represent the numerical fit functions in Eq.~\eqref{eq:deltaFit}.
In the right panel, the suppression scale $\Lambda$ is fixed
at $\Lambda = 3\times 10^{17}\,\textrm{GeV}$.
The scaling of $\delta_{\rm KG}$ with $\Lambda$ is trivial,
$\delta_{\rm KG} \propto \Lambda^{-2}$.
By definition, values of $\delta_{\rm F}$ larger than unity are unphysical;
see Eq.~\eqref{eq:ximax}.
For values of the Hubble rate greater than $H \simeq 8 \times 10^{13}\,\textrm{GeV}$,
the PLANCK constraint on the tensor-to-scalar ratio, $r \lesssim 0.11$, is violated.}
\label{fig:deltas}
\end{center}
\end{figure}

With Eq.~\eqref{eq:rhonum} at our disposal, we are now able to assess the relative
importance of the new terms in Eqs.~\eqref{eq:Friedmann} and \eqref{eq:KleinGordon}.
We are mainly interested in the following two ratios,
\begin{align}
\label{eq:deltaDef}
\delta_{\rm F} = \frac{\rho_{EE} + \rho_{BB}}{3 H^2 M_{\rm Pl}^2} \,, \quad
\delta_{\rm KG} =\left|\frac{\rho_{EB}/\Lambda}{3H \dot{a}}\right|
= \left|\frac{\rho_{EB}}{6\xi\Lambda^2H^2}\right| \,.
\end{align}
Here, $\delta_{\rm F}$ quantifies the hyper-EM contributions to the Friedmann
equation, while $\delta_{\rm KG}$ measures the importance of the source term
in the Klein-Gordon equation in comparison to the Hubble friction term.
For $4\lesssim \xi \lesssim 10$, these two ratios are well fit by the following
numerical expressions,
\begin{align}
\label{eq:deltaFit}
\delta_{\rm F} & \simeq 2.8 \times 10^{-4} \,
\exp\left[0.90 \times 2\pi\left(\xi-5\right)\right]
\left(\frac{H}{10^{13}\,\textrm{GeV}}\right)^2
\,, \\ \nonumber
\delta_{\rm KG} & \simeq 7.6 \times 10^{-4} \,
\exp\left[0.83 \times 2\pi\left(\xi-5\right)\right]
\left(\frac{H}{10^{13}\,\textrm{GeV}}\right)^2
\left(\frac{3\times10^{17}\,\textrm{GeV}}{\Lambda}\right)^2 \,.
\end{align}
These relations are the first important results of our analysis.
We stress that they represent numerical fit functions, which we obtain
by fitting $\delta_{\rm F}$ and $\delta_{\rm KG}$ as functions of
$e^{2\pi\xi}$, $H^2$, and $\Lambda^{-2}$.
The factors $0.90$ and $0.83$ in front of $2\pi\xi$ in Eq.~\eqref{eq:deltaFit}
account for the competition between the exponentials ($e^{2\pi\xi}$) and the inverse
powers ($\xi^{-3}$, $\xi^{-4}$, and $\xi^{-5}$) of $\xi$ in Eq.~\eqref{eq:rhoI}.
In Fig.~\ref{fig:deltas}, we compare our fit functions
with the corresponding exact expressions for $\delta_{\rm F}$ and $\delta_{\rm KG}$
in Eq.~\eqref{eq:deltaDef}.


From Eq.~\eqref{eq:deltaFit}, we see that the backreaction
from the gauge field on the inflationary dynamics is negligible,
at least for the chosen reference values.
This conclusion drastically changes as soon as we go to larger values of $\xi$ and $H$
as well as to smaller values of $\Lambda$.
Here, we find in particular an upper bound on $\xi$, such that the ratio $\delta_{\rm F}$
does not take values larger than unity; see Fig.~\ref{fig:deltas},
\begin{align}
\label{eq:ximax}
\delta_{\rm F} \leq 1 \quad\Rightarrow\quad
\xi \leq \xi_{\rm max} \left(H\right)
\simeq 6.4 - 0.82\, \log_{10}\left(\frac{H}{10^{13}\,\textrm{GeV}}\right) \,.
\end{align}
This bound is model-independent and needs to be obeyed by any model of pseudoscalar
inflation coupled to an Abelian gauge sector.
For $\xi$ values beyond this bound, one formally finds that more than 100\,\% of
the total energy density is stored in the hyper-EM field.
This signals that the backreaction from the excited
gauge fields is no longer negligible in the Friedmann equation; and hence the above
solutions are no longer trustable.
Meanwhile, the ratio $\delta_{\rm KG}$ can be varied independently,
even if $\xi$ satisfies Eq.~\eqref{eq:ximax}, simply by adjusting the
strength of the axion-gauge-field coupling.
According to Eq.~\eqref{eq:deltaFit}, lowering the suppression scale $\Lambda$ by a factor $10$
readily increases $\delta_{\rm KG}$ by two orders of magnitude.
For the same values of $\xi$ and $H$ as in Eq.~\eqref{eq:deltaFit},
$\xi = 5$ and $H = 10^{13}\,\textrm{GeV}$, but with
$\Lambda = 3 \times 10^{16}\,\textrm{GeV}$, the source term
in Eq.~\eqref{eq:KleinGordon} begins to compete with the Hubble friction term,
$\delta_{\rm KG} \sim 0.1$.
As we will see in the following, such small values of $\Lambda$, however,
turn out to be incompatible with the idea of baryogenesis from
pseudoscalar inflation.


\subsection{Hypermagnetic field at the end of inflation}
\label{subsec:initialcond}


As long as we stay sufficiently far away from the maximal $\xi$ value in
Eq.~\eqref{eq:ximax} and as long as the suppression scale $\Lambda$ is not chosen too small,
the effect of gauge field production merely represents a small
(and most often completely negligible) perturbation of the inflationary dynamics.
In this regime, we can therefore safely trust our analysis in the previous section.
In particular, we can use our result for the hypermagnetic field energy density, $\rho_{BB}$,
in Eq.~\eqref{eq:rhoI} to estimate the physical hypermagnetic field strength, $B_p$,
at any given time during inflation,
\begin{align}
\label{eq:BpRH}
B_p^2 = 2\,\rho_{BB} = \left<\bm{B}^2\right>
= \frac{1}{R^4} \int \frac{d^3\bm{k}}{\left(2\pi\right)^3} \, k^2 \left|A_+^k\right|^2
= 2\,\: \mathcal{I}_{BB}\left(\xi\right)\,\frac{e^{2\pi\xi}}{\xi^5}\, H^4 \,,
\end{align}
where we again neglect the vacuum contributions from the negative-helicity modes.
This field strength is the evident manifestation of \textit{primordial magnetogenesis}
in models of pseudoscalar inflation coupled to the hypercharge gauge field.
For typical values of $\xi$ and $H$, one finds
\begin{align}
\label{eq:bpRH}
B_p \simeq 1.1 \times 10^{49}\,\textrm{G}
\left(\frac{f_{BB}\left(\xi\right)}{f_{BB}\left(5\right)}\right)^{1/2}
\left(\frac{H}{10^{13}\,\textrm{GeV}}\right)^2 \,, \quad
f_{BB}\left(\xi\right) = \mathcal{I}_{BB}\left(\xi\right)\,\frac{e^{2\pi\xi}}{\xi^5} \,.
\end{align}
In the next section, we will discuss the postinflationary evolution of this
primordial hypermagnetic field, arguing that it is not completely erased
during the radiation-dominated era.
The primordial hypermagnetic field may, in fact, survive all the way up
to the present epoch and contribute to the intergalactic
magnetic fields that we observe today.


Another important quantity that characterizes the primordial hypermagnetic field is
the physical correlation length, $\lambda_p$.
To estimate $\lambda_p$, we compute the average
of all relevant wavelengths, weighted by their respective contributions
to the energy density $\rho_{BB}$,
\begin{align}
\label{eq:lambdapRH}
\lambda_p = \frac{1}{\rho_{BB}}\frac{1}{2R^4} \int \frac{d^3\bm{k}}{\left(2\pi\right)^3}
\frac{2\pi R}{k}\, k^2 \left|A_+^k\right|^2 =
\xi\,\frac{\mathcal{I}_\lambda\left(\xi\right)}{\mathcal{I}_{BB}\left(\xi\right)}\frac{2\pi}{H} \,.
\end{align}
Here, the integral function $\mathcal{I}_\lambda$ is defined
in analogy to the three functions in Eq.~\eqref{eq:integrals}
\begin{align}
\label{eq:Ilambda}
\mathcal{I}_{\lambda}\left(\xi\right) = \frac{\xi^4}{8\pi^2}\, e^{-\pi\,\xi}
\int_0^{x_{\rm UV}} dx\,x^2 \left|W_{\kappa_+,1/2}\left(-2ix\right)\right|^2 \,.
\end{align}
Similarly as the other integral functions, $\mathcal{I}_{\lambda}$ becomes
insensitive to the exact choice of $x_{\rm UV}$ as soon as it approaches
its asymptotic value.
For $\xi \gtrsim 4$, it is well approximated by
$\mathcal{I}_{\lambda} \simeq 8.7 \times 10^{-5}$;
see Fig.~\ref{fig:integrals}.
Together with the asymptotic value for $\mathcal{I}_{BB}$, this shows
that the hypermagnetic fields typically exhibit a correlation length that
extends over more than one Hubble radius,
\begin{align}
\lambda_p \simeq 3.0 \left(\frac{\xi}{5}\right) \lambda_H \,, \quad
\lambda_H  = \frac{2\pi}{H} \,.
\end{align}
More explicitly, we find that $\lambda_p$ typically takes values of
the following order of magnitude,
\begin{align}
\label{eq:lpRH}
\lambda_p \simeq 1.1 \times 10^{-50}\,\textrm{Mpc}
\left(\frac{f_\lambda\left(\xi\right)}{f_\lambda\left(5\right)}\right)
\left(\frac{10^{13}\,\textrm{GeV}}{H}\right) \,, \quad
f_\lambda\left(\xi\right) =
\xi\,\frac{\mathcal{I}_\lambda\left(\xi\right)}{\mathcal{I}_{BB}\left(\xi\right)}\,.
\end{align}


The above expressions for $B_p$ and $\lambda_p$ in Eqs.~\eqref{eq:BpRH}
and \eqref{eq:lambdapRH} are valid at any time during inflation. %
In the following, we are however going to be mostly interested in the values
of $B_p$ and $\lambda_p$ at the end of inflation, i.e., at the onset of
reheating.
In this paper, we will work in the approximation of instant reheating,
such that the end of inflation coincides with the beginning of the
radiation-dominated era.
To find the values of $B_p$ and $\lambda_p$ at this time, it is, therefore,
sufficient to simply evaluate Eqs.~\eqref{eq:BpRH} and \eqref{eq:lambdapRH}
for $H = H_{\rm rh}$ and $\xi = \xi_{\rm rh}$, where $H_{\rm rh}$ and $\xi_{\rm rh}$
respectively denote the Hubble rate and the instability parameter at the end of inflation.
Both quantities are model-dependent, which is why we will treat
them as free parameters in the following.
At this point it is interesting to note that, for most models of interest,
$\xi_{\rm rh}$ is entirely controlled by the strength of the axion-gauge-field coupling.
To see this, let us suppose that the end of inflation is triggered by a
violation of the first slow-roll condition.
That is, inflation ends because the Hubble parameter $H$ is no longer quasi-constant.
This condition is conveniently quantified in terms of the slow-roll parameter $\varepsilon$.
Let us assume for now that the backreaction from gauge field production is negligible.
In the usual slow-roll approximation, one then has
\begin{align}
\varepsilon = \frac{d\,\ln H}{d N_e} \approx
\frac{M_{\rm Pl}^2}{2} \left(\frac{d\,\ln V}{da}\right)^2 \approx
\frac{\dot{a}^2}{2\,H^2 M_{\rm Pl}^2} \,,
\end{align}
where $N_e$ denotes the number of e-folds until the end of inflation.
Next, let us rewrite the condition $\varepsilon \sim 1$ at the end of inflation
in terms of $\xi_{\rm rh}$ and $\Lambda$.
This yields
\begin{align}
\label{eq:xiRH}
\varepsilon \approx \frac{2\,\xi_{\rm rh}^2\Lambda^2}{M_{\rm Pl}^2} \sim 1
\quad\Rightarrow\quad \xi_{\rm rh} \sim \frac{M_{\rm Pl}}{\sqrt{2}\,\Lambda}
\simeq 5.7 \left(\frac{3\times10^{17}\,\textrm{GeV}}{\Lambda}\right) \,.
\end{align}
Together with Eq.~\eqref{eq:deltaFit}, this result confirms that, for
$\Lambda \gtrsim 3 \times 10^{17}\,\textrm{GeV}$,
the backreaction on the inflationary dynamics is mostly negligible at all times.
For smaller values of $\Lambda$, the ratio $\delta_{\rm KG}$ however quickly approaches
values of order unity towards the end of inflation.


\section{Gauge field evolution after inflation}
\label{sec:evolution}


We now turn to the description of the postinflationary evolution of the
primordial gauge fields.
We will discuss in turn the different stages until the beginning
of the inverse cascade regime (see Sec.~\ref{subsec:untilIC}), until the electroweak
phase transition (see Sec.~\ref{subsec:untilEWPT}),
and until today (see Sec.~\ref{subsec:untiltoday}).


\subsection{From the end of inflation to the onset of the inverse cascade regime}
\label{subsec:untilIC}


As stressed several times before, we are going to work in the approximation
of instant reheating.%
\footnote{Similarly, we also assume that there is no charged plasma even as
a subdominant component of the universe until the end of inflation.
The presence of such a charged plasma component already during the stage of
inflation might prevent the hypermagnetic helicity from developing and, hence, change our estimate.}
That is, we make the simplifying assumption that, at the end of inflation,
the vacuum energy density driving inflation is converted instantaneously
into thermal radiation,
\begin{align}
\label{eq:instantRH}
\rho_{\rm inf}\left(t_{\rm rh}\right) = 3\, H_{\rm rh}^2\, M_{\rm Pl}^2
\quad\rightarrow\quad
\rho_{\rm rad}\left(t_{\rm rh}\right) = \frac{\pi^2}{30}\,g_*\, T_{\rm rh}^4 \,,
\end{align}
where $g_* = 106.75$ denotes the effective number of relativistic DOFs
in the standard model.
We consequently neglect the period of inflaton oscillations after inflation
as well as the gradual production of (charged) particles in inflaton decays.
This assumption simplifies our analysis considerably\,---\,given the fact that
the charged particles in the emerging plasma actually interfere with the evolution
of the primordial gauge fields.%
\footnote{The oscillations of the inflaton field would enhance the production of
primordial gauge fields~\cite{Adshead:2016iae}.
But, at the same time, the high electric conductivity of the charged plasma
would suppress the hypermagnetic helicity~\cite{Fujita:2015iga}.}
A reliable description of this complicated process however requires a dedicated
numerical simulation that takes into account both nonperturbative particle production and MHD,
which is not yet available and which is certainly beyond the scope of this work.
The assumption of instant reheating moreover allows us to eliminate the reheating
temperature $T_{\rm rh}$ as a free parameter in our scenario.
According to Eq.~\eqref{eq:instantRH}, we can simply express $T_{\rm rh}$
in terms of the Hubble rate at the end of inflation, $H_{\rm rh}$,
\begin{align}
T_{\rm rh} = \sqrt{M_*\,H_{\rm rh}} \simeq 2.7 \times 10^{15}\,\textrm{GeV}
\left(\frac{H_{\rm rh}}{10^{13}\,\textrm{GeV}}\right)^{1/2} \,, \quad
M_* = \left(\frac{90}{\pi^2\,g_*}\right)^{1/2} M_{\rm Pl} \,.
\end{align}
By employing this relation, we choose to discard all details of the reheating process.
While this introduces an uncertainty to some degree,
it also makes our analysis more model-independent.

\begin{figure}
\begin{center}
\includegraphics[width=0.75\textwidth]{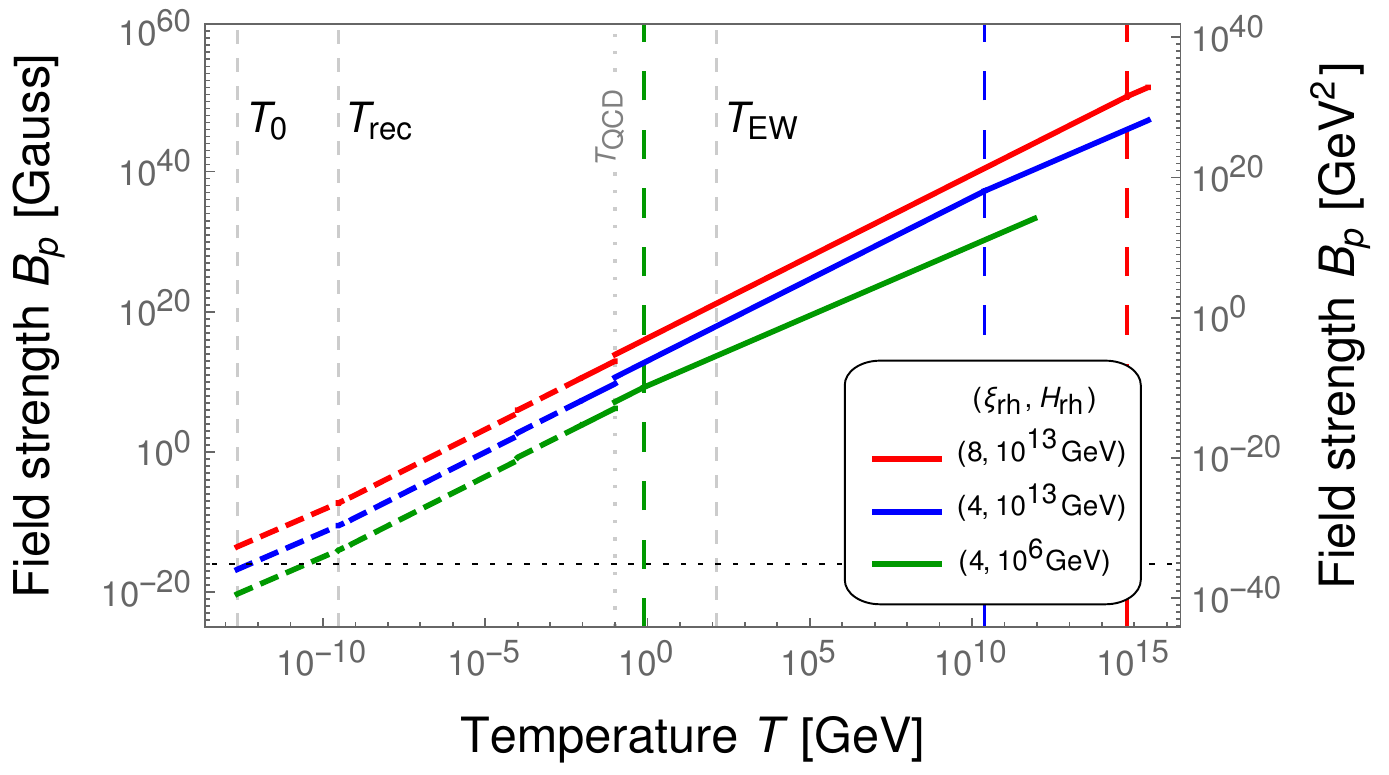}

\bigskip
\includegraphics[width=0.75\textwidth]{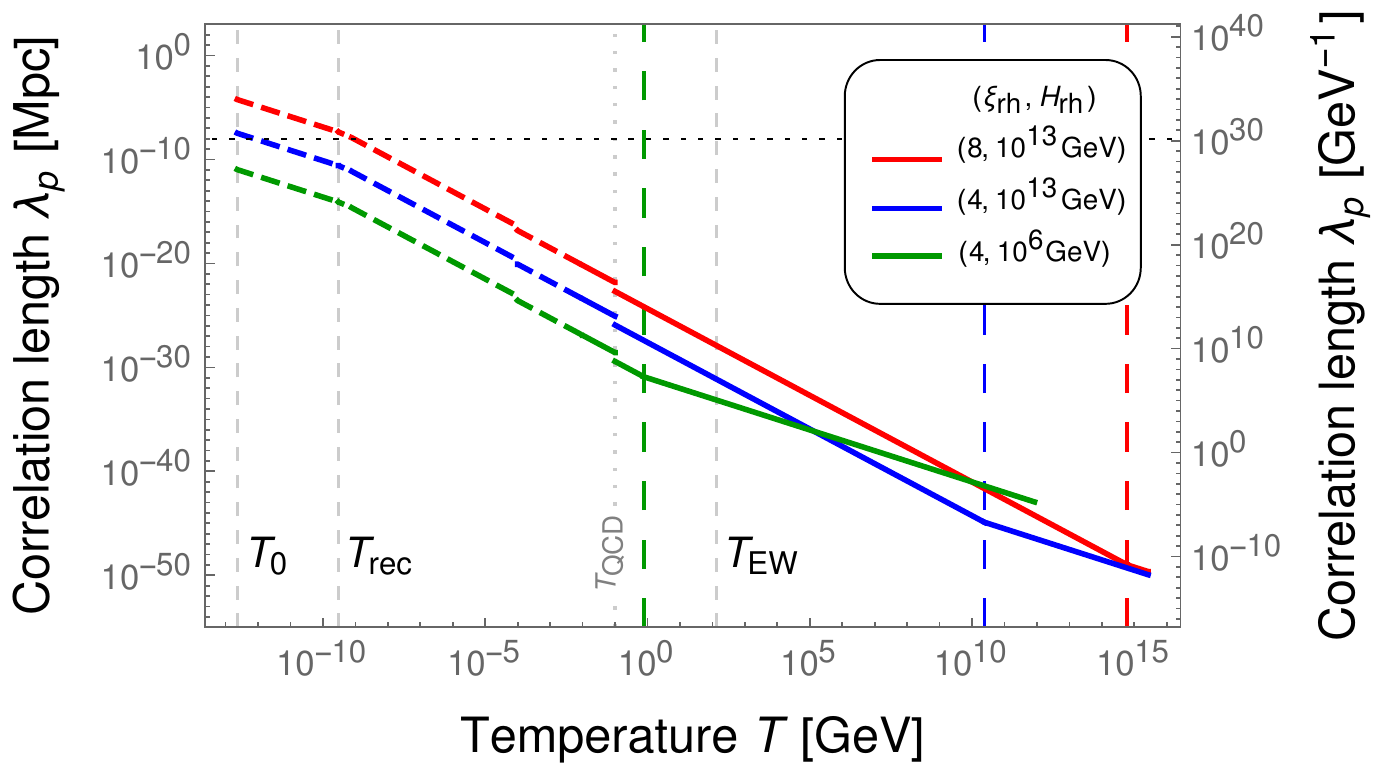}
\caption{Physical field strength $B_p$ \textbf{(upper panel)} and physical
correlation length $\lambda_p$ \textbf{(lower panel)} of the hypermagnetic
$\bm{B}$ field as functions of the radiation temperature $T$ for representative
values of $H$ and $\xi$ at the end of inflation.
The vertical dotted lines mark the respective temperatures at which the
adiabatic regime transitions into the inverse cascade regime.
Both plots account for the decrease in the effective number of DOFs
in the course of the expansion.
The kinks around $T \sim 100\,\textrm{MeV}$ correspond, e.g., to the
QCD phase transition.
For $T<10\,\textrm{MeV}$, damping effects
might become important (see \cite{Durrer:2013pga,Kahniashvili:2012uj,Banerjee:2004df})
and our description of the magnetic field evolution becomes less accurate.
For this reason, we only draw dashed lines in the low-temperature regime.}
\label{fig:evolution}
\end{center}
\end{figure}

To describe the behavior of the primordial gauge fields
after reheating, we shall follow the discussion
in~\cite{Durrer:2013pga,Kahniashvili:2012uj,Banerjee:2004df}
(see also \cite{Widrow:2011hs,Kandus:2010nw,Brandenburg:2004jv}).
Our first observation is that, once the plasma is in place, the hyper-EM field
begins to interact with hypercharged particles in the thermal bath.
This interaction makes the primordial hyperelectric fields vanishingly small,
$\bm{E} \simeq 0$ (i.e., $\bm{E}$ becomes suppressed by the large electric conductivity),
leaving us mainly with the hypermagnetic $\bm{B}$ field.
In the following, we will assume that, initially, the backreaction from the charged
particles has neither an impact on the overall strength of the
hypermagnetic field, $B_p$, nor on its correlation length, $\lambda_p$.
The starting point of our analysis are, therefore, our results for $B_p$
and $\lambda_p$ that we obtained in
Sec.~\ref{subsec:initialcond}; see Eqs.~\eqref{eq:BpRH} and \eqref{eq:lambdapRH},
\begin{align}
\label{eq:initialcond}
B_p^{\rm rh} = \left(2\,\mathcal{I}_{BB}\right)^{1/2}
\frac{e^{\pi\xi_{\rm rh}}}{\xi_{\rm rh}^{5/2}}\, H_{\rm rh}^2 \simeq
1.7 \times 10^{-2}\,\frac{e^{\pi\xi_{\rm rh}}}{\xi_{\rm rh}^{5/2}}\, H_{\rm rh}^2
\,, \quad
\lambda_p^{\rm rh} =
\xi_{\rm rh}\,\frac{\mathcal{I}_\lambda}{\mathcal{I}_{BB}}\frac{2\pi}{H_{\rm rh}} \simeq
3.7\,\frac{\xi_{\rm rh}}{H_{\rm rh}} \,.
\end{align}
We are now going to outline how these two quantities behave as functions
of the radiation temperature $T$, as the universe expands.
Our final results are summarized schematically in Fig.~\ref{fig:evolution},
which illustrates the time dependence of $B_p$ and $\lambda_p$
for different values of $\xi_{\rm rh}$ and $H_{\rm rh}$.


At early times, i.e., directly after reheating, we expect that both the field strength $B_p$
as well as the correlation length $\lambda_p$ simply redshift adiabatically~\cite{Fujita:2016igl},
\begin{align}
\label{eq:adiabaticredshift}
B_p\left(T\right) = \left(\frac{R_{\rm rh}}{R\left(T\right)}\right)^2 B_p^{\rm rh} \,, \quad
\lambda_p\left(T\right) = \left(\frac{R\left(T\right)}{R_{\rm rh}}\right) \lambda_p^{\rm rh} \,.
\end{align}
This expectation is justified by the fact that, initially, the correlation length $\lambda_p$
is much longer than the eddy scale of the
velocity fields of the charged plasma, $\lambda_T \simeq v t$,
(see also the discussion in the next section) implying that
the charged plasma cannot affect the evolution of the hypermagnetic field.
During radiation domination and for a constant number of effective DOFs,
the scale factor $R$ increases in inverse proportion to the plasma temperature,
$R \propto 1/T$.
During the early phase of adiabatic expansion, $B_p$ therefore
drops like $T^2$, while $\lambda_p$ grows like $1/T$.


\subsection{From the onset of the inverse cascade regime to the electroweak crossover}
\label{subsec:untilEWPT}


In the course of the further evolution, the interaction of the $\bm{B}$ field
and the charged plasma (described by the velocity field ${\bm v}$)
results in a complicated co-evolution of both fields, governed by the MHD equations:
The $\bm{B}$ field induces a ${\bm v}$ field and
the $\bm{v}$ field back-reacts on the evolution of the ${\bm B}$ field, which likely results
in turbulent field configurations.
If the charged plasma develops a turbulence, the scale up to which the velocity field
is capable of affecting the $\bm{B}$ field can be estimated in terms
of the turbulence (or eddy) scale $\lambda_T$,
\begin{align}
\label{eq:lambdaT}
\lambda_T \simeq v\, t = \frac{v}{2H} \,, \quad v = \left|{\bm v}\right| \,.
\end{align}
As long as $\lambda_T \ll \lambda_p$,
the $\bm{v}$ field affects the $\bm{B}$ field
only on small scales and the evolution of $B_p$ and $\lambda_p$ remains unaffected.
Both the turbulence scale $\lambda_T$ and the correlation length $\lambda_p$ grow with time.
However, $\lambda_T$ grows faster than $\lambda_p$,
such that, after some finite time, the turbulence scale catches up with
the correlation length, $\lambda_T \sim \lambda_p$.
After that, the ${\bm B}$ field can no longer evolve adiabatically.
Indeed, it has been observed in MHD simulations that a maximally helical magnetic
field generates a turbulent plasma  and that the kinetic energy of the plasma waves
becomes comparable to (or equilibrated with) the energy stored
in the hypermagnetic field~\cite{Kahniashvili:2012uj,Banerjee:2004df},
$\rho_{\rm kin} \sim \rho_{BB}$.
This means that the amplitude of the $\bm{v}$ field is comparable to the
Alfv\'en velocity $v_A$.
In the nonrelativistic limit, $v_A \ll 1$,
the Alfv\'en velocity is given as~\cite{Gedalin:1993aa},
\begin{align}
\label{eq:vA}
v \sim v_A = \frac{v_A^0}{\sqrt{1+\left(v_A^0\right)^2}} \sim v_A^0 \,, \quad
v_A^0 = \frac{B_p}{\sqrt{\rho_{\rm ch} + p_{\rm ch}}} \,, \quad
\rho_{\rm ch} =  \frac{\pi^2}{30}\,g_{*,\rm ch} \, T^4 \,, \quad
p_{\rm ch} = \frac{\rho_{\rm ch}}{3} \,,
\end{align}
with $\rho_{\rm ch}$ and $p_{\rm ch}$ denoting the energy density and pressure
of the hypercharged particles in the plasma.
$g_{*,\rm ch} = 82.75$ counts the effective number of relativistic DOFs carrying
nonzero hypercharge in the standard model.
In the following, we will not distinguish between $v_A$ and $v_A^0$
and simply approximate $v_A \approx v_A^0$.
Combining Eqs.~\eqref{eq:lambdaT} and \eqref{eq:vA} and
assuming that the ${\bm B}$ and $\bm{v}$ fields are equilibrated
even in the adiabatic regime, we find for $\lambda_T$ in the adiabatic regime
\begin{align}
\lambda_T \propto \frac{B_p}{\sqrt{\rho_{\rm ch}}\,H} \propto R^2 \propto \frac{1}{T^2} \,
\end{align}
Indeed, this corresponds to a faster growth than in the case of $\lambda_p$, which simply
scales like $1/T$.


Once the turbulence scale has caught up with the correlation length,
$\lambda_T \sim \lambda_p$, the hypermagnetic field enters into the
inverse cascade regime~\cite{Frisch:1975aa,Pouquet:1976zz,Kahniashvili:2012uj}.
From this point on, the growth of $\lambda_p$ is simply driven by the turbulence
scale $\lambda_T$, such that $\lambda_T \sim \lambda_p$ at all subsequent times,
\begin{align}
\label{eq:ICrel1}
\lambda_p \sim \lambda_T \propto \frac{B_p}{\sqrt{\rho_{\rm ch}}\,H}
\quad\Rightarrow\quad
\frac{\lambda_p}{B_p} \propto \frac{1}{\sqrt{\rho_{\rm ch}}\,H} \propto R^4
\propto \frac{1}{T^4} \,.
\end{align}
This relation is, however, not yet sufficient to fully estimate
the scaling behavior of $B_p$ and $\lambda_p$ during the inverse cascade regime.
In addition to Eq.~\eqref{eq:ICrel1}, we need a second, independent
relation between $\lambda_T$ and $B_p$.
At this point, it comes in handy that, as a consequence of
the high hyperelectric conductivity of the charged plasma,
the comoving helicity density $h_c$ is approximately
conserved at high temperatures; see, e.g., \cite{Anber:2015yca,Fujita:2016igl}
and references therein,%
\footnote{Based on Amp\`ere's and Ohm's laws, one can show that the
time derivative of $h_c$ is suppressed by the inverse of the hyperelectric
conductivity, $\dot{h}_c \propto 1/\sigma$.
The fact that $h_c$ is conserved to good approximation is, therefore, a direct consequence
of the large (but finite) conductivity of the standard model plasma,
$\sigma \sim 10^2\,T$~\cite{Baym:1997gq,Arnold:2000dr}.
\label{fn:hcdot}}
\begin{align}
\label{eq:hc}
h_c = \lim_{V\rightarrow \infty}\frac{1}{V}
\int_V d^3 \bm{x}\: \bm{A}_c\cdot\bm{B}_c \sim \textrm{const} \,.
\end{align}
Here, the integral over the volume $V$ represents nothing but a spatial average,
$h_c = \left<\bm{A}_c\bm{B}_c\right>$.
Moreover, we emphasize that both vector fields, $\bm{A}_c \equiv \bm{A}$
and $\bm{B}_c = R^2 \bm{B}$, correspond to comoving quantities.
We roughly estimate the typical size of $\bm{A}_c$ as
$A_c \sim \lambda_c/\left(2\pi\right) B_c \propto R\, \lambda_p B_p$, such that
\begin{align}
\label{eq:ICrel2}
h_c = R^2 \left<\bm{A}\bm{B}\right> \propto R^3\lambda_pB_p^2\sim \textrm{const} \,.
\end{align}
Together with Eq.~\eqref{eq:ICrel1}, this relation then yields
the scaling behavior of $B_p$ and $\lambda_p$,
\begin{align}
\label{eq:ICredshift}
B_p \propto \frac{1}{R^{7/3}} \propto T^{7/3} \,, \quad
\lambda_p \propto R^{5/3} \propto \frac{1}{T^{5/3}} \,,
\end{align}
which coincides with the scaling laws of the inverse cascade found in
MHD simulations~\cite{Kahniashvili:2012uj,Banerjee:2004df}.


We stress that all of the relations in Eqs.~\eqref{eq:lambdaT}, \eqref{eq:vA},
\eqref{eq:ICrel1}, and \eqref{eq:ICrel2} are simply rough estimates.
A more careful treatment would require a full-fledged MHD
simulation~\cite{Durrer:2013pga,Brandenburg:2004jv}, which is beyond the scope of this work.
Moreover, the study of primordial magnetic fields in MHD
simulations is still the subject of on-going work in the literature.
In anticipation of new simulations, we shall therefore
settle for the estimates above, leaving any refinement of our analysis
for future work.


Next, let us determine the temperature at the onset
of the inverse cascade regime.
We find the transition temperature, $T = T_{\rm ic}$, simply
by solving the condition $\lambda_T\left(T_{\rm ic}\right) = \lambda_p\left(T_{\rm ic}\right)$
for $T_{\rm ic}$,
\begin{align}
\label{eq:TIC}
\frac{T_{\rm ic}}{T_{\rm rh}} =
\frac{\mathcal{I}_{BB}^{3/2}}{4\sqrt{2}\,\pi\,\mathcal{I}_\lambda}
\left(\frac{g_*}{g_{*,\rm ch}}\right)^{1/2}
\frac{e^{\pi \xi_{\rm rh}}}{\xi_{\rm rh}^{7/2}}
\frac{H_{\rm rh}}{M_{\rm Pl}} \simeq
1.3 \times 10^{-3}\,\frac{e^{\pi \xi_{\rm rh}}}{\xi_{\rm rh}^{7/2}}
\frac{H_{\rm rh}}{M_{\rm Pl}} \,,
\end{align}
where we used that $T_{\rm rh} = \sqrt{M_* H_{\rm rh}}$.
The Alfv\'en velocity $v_A$ at this temperature is given as
\begin{align}
v_A =
\frac{\mathcal{I}_{BB}^{1/2}}{\sqrt{2}}
\left(\frac{g_*}{g_{*,\rm ch}}\right)^{1/2}
\frac{e^{\pi \xi_{\rm rh}}}{\xi_{\rm rh}^{5/2}}
\frac{H_{\rm rh}}{M_{\rm Pl}} \simeq
9.7 \times 10^{-3}\,\frac{e^{\pi \xi_{\rm rh}}}{\xi_{\rm rh}^{5/2}}
\frac{H_{\rm rh}}{M_{\rm Pl}} \,,
\end{align}
Furthermore, we are now in the position to calculate the field strength
as well as the correlation length of the hypermagnetic field at $T = T_{\rm ic}$.
Combining Eqs.~\eqref{eq:initialcond}, \eqref{eq:adiabaticredshift}, and \eqref{eq:TIC},
we obtain
\begin{align}
B_p^{\rm ic} & = \left(\frac{T_{\rm ic}}{T_{\rm rh}}\right)^2 B_p^{\rm rh} =
\frac{\mathcal{I}_{BB}^{7/2}}{16\sqrt{2}\,\pi^2\,\mathcal{I}_\lambda^2}
\frac{g_*}{g_{*,\rm ch}}
\frac{e^{3\pi \xi_{\rm rh}}}{\xi_{\rm rh}^{19/2}}
\frac{H_{\rm rh}^4}{M_{\rm Pl}^2} \simeq
2.9 \times 10^{-8}\,\frac{e^{3\pi \xi_{\rm rh}}}{\xi_{\rm rh}^{19/2}}
\frac{H_{\rm rh}^4}{M_{\rm Pl}^2} \,,
\\ \nonumber
\lambda_p^{\rm ic} & = \left(\frac{T_{\rm rh}}{T_{\rm ic}}\right) \lambda_p^{\rm rh} =
\frac{8\sqrt{2}\,\pi^2\,\mathcal{I}_\lambda^2}{\mathcal{I}_{BB}^{5/2}}
\left(\frac{g_{*,\rm ch}}{g_*}\right)^{1/2}
\frac{\xi_{\rm rh}^{9/2}}{e^{\pi \xi_{\rm rh}}}
\frac{M_{\rm Pl}}{H_{\rm rh}^2} \simeq
2.9 \times 10^3\,\frac{\xi_{\rm rh}^{9/2}}{e^{\pi \xi_{\rm rh}}}
\frac{M_{\rm Pl}}{H_{\rm rh}^2} \,.
\end{align}
At temperatures below $T_{\rm ic}$, the field strength $B_p$ behaves as follows;
see Eq.~\eqref{eq:ICredshift},
\begin{align}
\label{eq:BpIC}
B_p\left(T\right) & = \left(\frac{T}{T_{\rm ic}}\right)^{7/3} B_p^{\rm ic} =
\left(\frac{T}{T_{\rm rh}}\right)^{7/3}\left[
16\,\pi\,\mathcal{I}_\lambda\left(\frac{g_{*,\rm ch}}{g_*}\right)^{1/2}
\frac{e^{2\pi \xi_{\rm rh}}}{\xi_{\rm rh}^4}
H_{\rm rh}^5 M_{\rm Pl}\right]^{1/3}
\\ \nonumber
& \simeq 0.16
\left(\frac{T}{T_{\rm rh}}\right)^{7/3}\left(\frac{e^{2\pi \xi_{\rm rh}}}{\xi_{\rm rh}^4}
H_{\rm rh}^5 M_{\rm Pl}\right)^{1/3} \,,
\end{align}
whereas for the correlation length $\lambda_p$, we find
\begin{align}
\label{eq:lpIC}
\lambda_p\left(T\right) & = \left(\frac{T_{\rm ic}}{T}\right)^{5/3} \lambda_p^{\rm ic} =
\left(\frac{T_{\rm rh}}{T}\right)^{5/3}\left(
\frac{\pi\,\mathcal{I}_\lambda}{4}
\frac{g_*}{g_{*,\rm ch}}
\frac{e^{2\pi \xi_{\rm rh}}}{\xi_{\rm rh}^4}
\frac{1}{H_{\rm rh} M_{\rm Pl}^2}\right)^{1/3}
\\ \nonumber
& \simeq 4.5 \times 10^{-2}
\left(\frac{T_{\rm rh}}{T}\right)^{5/3}
\left(\frac{e^{2\pi \xi_{\rm rh}}}{\xi_{\rm rh}^4}
\frac{1}{H_{\rm rh} M_{\rm Pl}^2}\right)^{1/3} \,.
\end{align}
Note that we assumed a constant effective number of DOFs in both
Eq.~\eqref{eq:BpIC} and Eq.~\eqref{eq:lpIC}.


\subsection{From the electroweak crossover to the present epoch}
\label{subsec:untiltoday}


\begin{figure}
\begin{center}
\includegraphics[width=0.48\textwidth]{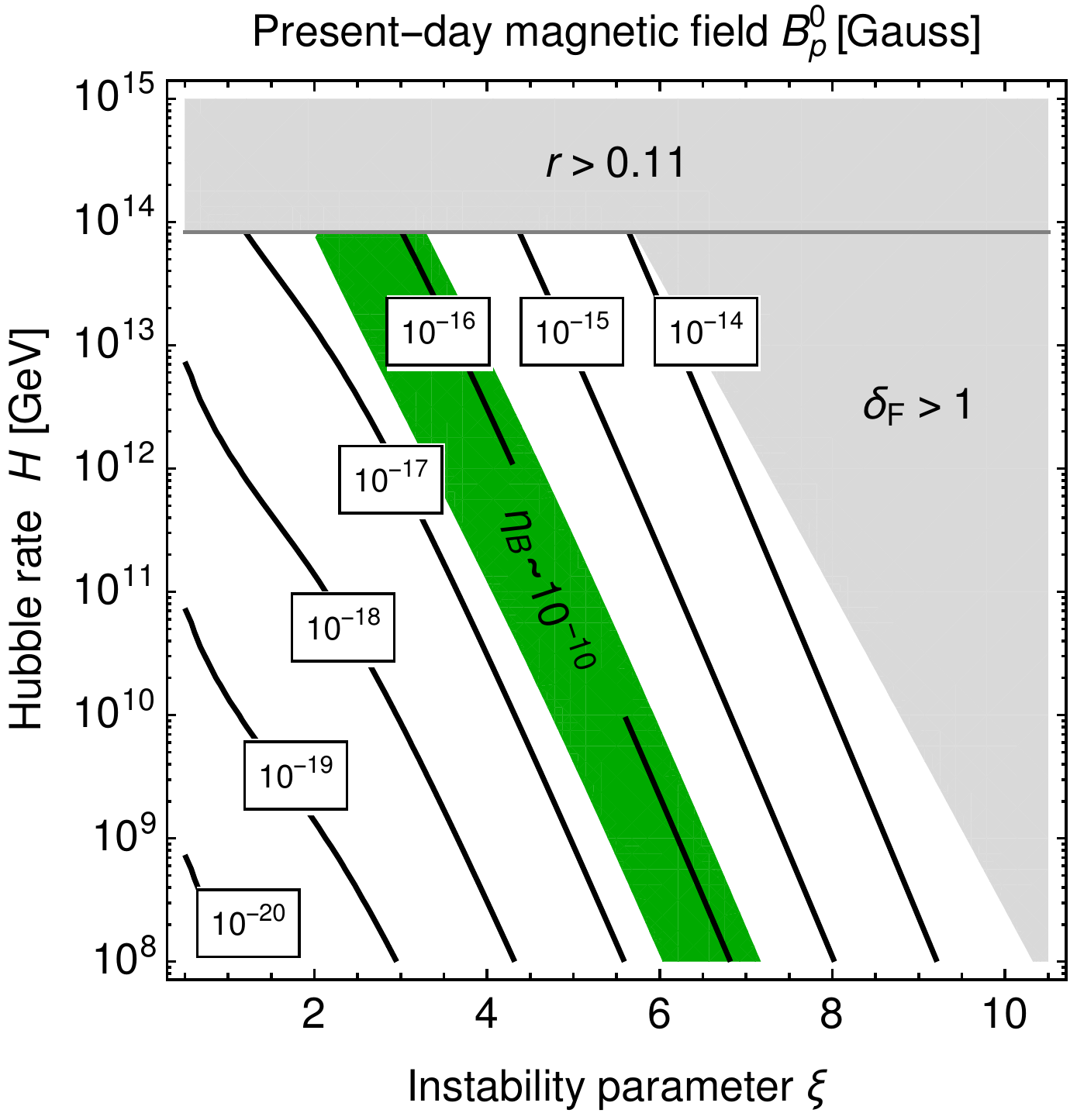}
\caption{Present-day strength of the physical magnetic field, $B_p^0$,
as a function of the instability parameter $\xi$ and the Hubble rate $H$;
see Eq.~\eqref{eq:Bp0}.
Here, both $\xi$ and $H$ are understood to correspond to the respective
values at the end of inflation, $\xi \equiv \xi_{\rm rh}$ and $H \equiv H_{\rm rh}$.
The green band illustrates the region in parameter space where baryogenesis
around the time of EWSB results in a baryon asymmetry $\eta_B$
in accord with the observed value, $\eta_B^{\rm obs} \sim 10^{-10}$;
see Eq.~\eqref{eq:etaB} and Fig.~\ref{fig:etaB}.
The gray-shaded regions are the same as in Fig.~\ref{fig:deltas}.}
\label{fig:Bp0}
\end{center}
\end{figure}


The evolution of the field strength and correlation length at late times can
be described by standard techniques with the assumption that
the magnetic fields evolve according to the inverse cascade until recombination
and evolve adiabatically again after that until today.
In the usual $\Lambda$CDM model (without any additional stages
of late-time entropy production or the like), we can readily relate
the values of $B_p$ and $\lambda_p$ around the time of
EWSB to their values in the present epoch,
\begin{align}
B_p^0 & \simeq 1.1\times 10^{-14}\,\textrm{G}\: \bigg(\frac{B_p^{\rm ew}}{10^{20}\,\textrm{G}}\bigg)
\left(\frac{100\,\textrm{GeV}}{T_{\rm ew}}\right)^{7/3} \,,
\\ \nonumber
\lambda_p^0 & \simeq 0.40\,\textrm{pc}\:
\bigg(\frac{\lambda_p^{\rm ew}}{10^{-29}\,\textrm{Mpc}}\bigg)
\left(\frac{T_{\rm ew}}{100\,\textrm{GeV}}\right)^{5/3} \,,
\end{align}
which is consistent with the corresponding relations in \cite{Fujita:2016igl,Kamada:2016eeb}.
Here, the values of $B_p^{\rm ew}$ and $\lambda_p^{\rm ew}$
simply follow from evaluating our results in
Eq.~\eqref{eq:BpIC} and Eq.~\eqref{eq:lpIC} at $T = T_{\rm ew} \sim 100\,\textrm{GeV}$, i.e.,
the temperature at the time of EWSB.%
\footnote{During EWSB, the hypermagnetic ${\bm B}_Y$ field
turns into the electromagnetic ${\bm B}_{\rm EM}$ field.
The amplitudes $\left|\bm{B}_Y\right|$ and $\left|\bm{B}_{\rm EM}\right|$
are, however, continuously connected~\cite{Kamada:2016cnb},
which is why we do not distinguish between them here.\smallskip}
We then obtain the following final expression for
the present-day strength of the physical magnetic field,
\begin{align}
\label{eq:Bp0}
B_p^0 & \simeq 6.0 \times 10^{-18}\,\textrm{G}
\left[\mathcal{I}_\lambda \left(\frac{g_{*,\rm ch}}{g_*}\right)^{1/2}
\frac{e^{2\pi\xi_{\rm rh}}}{\xi_{\rm rh}^4}\right]^{1/3}
\left(\frac{H_{\rm rh}}{10^{13}\,\textrm{GeV}}\right)^{1/2}
\\ \nonumber
& \simeq 2.5 \times 10^{-19}\,\textrm{G}
\left(\frac{e^{2\pi\xi_{\rm rh}}}{\xi_{\rm rh}^4}\right)^{1/3}
\left(\frac{H_{\rm rh}}{10^{13}\,\textrm{GeV}}\right)^{1/2} \,,
\end{align}
which we plot as a function of $\xi_{\rm rh}$ and $H_{\rm rh}$ in Fig.~\ref{fig:Bp0}.
This relation illustrates how the explosive production of gauge fields during pseudoscalar
inflation results in magnetic fields on astrophysical scales in the present epoch.
Note that $B_p^0$ in Eq.~\eqref{eq:Bp0} does not depend on the exact
value of $T_{\rm ew}$.
Moreover, it only depends on $\mathcal{I}_\lambda$ and
is independent of the integral function $\mathcal{I}_{BB}$.
Meanwhile, we find that the present-day value of the correlation length, $\lambda_p^0$,
satisfies exactly the relation which one expects for causally
generated magnetic fields~\cite{Banerjee:2004df}; see also
Eq.~\eqref{eq:benchmark} and footnote~\ref{fn:units},
\begin{align}
\label{eq:lp0}
\lambda_p^0 \simeq 0.28 \,\textrm{pc} \bigg(\frac{B_p^0}{10^{-14}\,\textrm{G}}\bigg)
\simeq \frac{1.0 \,\textrm{pc}}{\left(4\pi\right)^{1/2}}
\bigg(\frac{B_p^0}{10^{-14}\,\textrm{G}}\bigg)  \,.
\end{align}
We stress that this result is based on the strict relation $v=v_A$; see Eq.~\eqref{eq:vA}.
However, there are also MHD simulations suggesting that $v$ might in fact be slightly
suppressed compared to the Alfv\'en velocity,
$v \simeq \mathcal{O}\left(0.1\right) v_A$~\cite{Kahniashvili:2012uj,Banerjee:2004df}.
Thus, our above estimates come with at least an $\mathcal{O}\left(10\right)$ uncertainty.%
\footnote{Note that we also omitted possible
damping effects at low temperatures, $T< 10\,\textrm{MeV}$,
due to processes such as neutrino and photon free streaming.
However, despite these effects, it has been demonstrated that both the field
strength and the correlation length eventually reach the same values
as in the simple inverse-cascade estimate~\cite{Banerjee:2004df}.}
Nonetheless, we expect our expressions to catch the basic \textit{qualitative} features of the
magnetic field from pseudoscalar inflation, in particular, the relation
between the inflationary parameters $H$ and $\xi$ on the one hand and the
quantities $B_p$ and $\lambda_p$ on the other hand.


\section{Implications for baryon asymmetry and gravitational waves}
\label{sec:BAUGWs}


The primordial gauge fields generated during inflation have important
phenomenological consequences.
Not only do they seed the intergalactic magnetic fields that permeate
our Universe today (see Eq.~\eqref{eq:Bp0}), they also lead to
the generation of a nonzero baryon number around the time of EWSB (see Sec.~\ref{subsec:BAU})
as well as to a signal in the stochastic GW background at high frequencies
(see Sec.~\ref{subsec:GWs}).
We shall now discuss these two phenomena in turn.


\subsection{Baryogenesis from pseudoscalar inflation}
\label{subsec:BAU}


The gauge fields generated during inflation are maximally helical.
This can be seen explicitly from our analysis in Sec.~\ref{subsec:EOMs},
where we showed that only modes in one helicity eigenstate
are exponentially amplified during inflation, while the other helicity
modes stay at the vacuum level; see Eq.~\eqref{eq:Asolution}.
Moreover, we found that the sign of the final helicity depends on the sign of
the inflaton velocity, $\textrm{sgn}\,\mathcal{H} = \textrm{sgn}\,\dot{a}$.
In Sec.~\ref{subsec:EOMs}, we chose $\dot{a} > 0$, in order to achieve
positive helicity.


Changes in the comoving helicity density $h_c$ (see Eq.~\eqref{eq:hc}) after
inflation are suppressed by the hyperelectric conductivity of the thermal plasma,
$\dot{h}_c \propto 1/\sigma$; see footnote~\ref{fn:hcdot}.
Therefore, given the large value of $\sigma$ in the standard model,
$\sigma \sim 10^2\,T$~\cite{Baym:1997gq,Arnold:2000dr}, $h_c$
is \textit{approximately} conserved after inflation; see Eq.~\eqref{eq:ICrel2}.
At the same time, it is important to remember that any change in the
hypermagnetic helicity results in the production of baryon number $B$
and lepton number $L$.
This is reflected in Eq.~\eqref{eq:anomaly},
which follows from the chiral triangle anomaly in the standard model.
Therefore, even slight changes in $h_c$, because of the finite
conductivity $\sigma$, are physically relevant
as soon as we turn our attention to the time evolution of $B$ and $L$.
This observation is the basis for the scenario of \textit{baryogenesis via decaying hypermagnetic
helicity}~\cite{Anber:2015yca,Cado:2016kdp,Kamada:2016cnb,Kamada:2016eeb,Fujita:2016igl}.
In the following, we will illustrate how this scenario fits together
with our analysis of primordial magnetogenesis in models of pseudoscalar
inflation.
In doing so, we will follow the discussion in~\cite{Kamada:2016cnb}
(see also \cite{Kamada:2016eeb}).


In analogy to $h_c$ in Eq.~\eqref{eq:hc}, we may define the physical
helicity density $h_p$ as follows,
\begin{align}
\label{eq:hp}
h_p = \frac{h_c}{R^3} = \lim_{V\rightarrow \infty}\frac{1}{V}
\int_V d^3 \bm{x}\: \bm{A}_p\cdot\bm{B}_p = \left<\bm{A}_p\bm{B}_p\right> \,.
\end{align}
Here and only here, $\bm{A}_p$ is defined as $\bm{A}_p = \bm{A}/R$, whereas
$\bm{B_p} \equiv \bm{B}$ is nothing but the ordinary physical hypermagnetic $\bm{B}$
field.
Again, the angle brackets in Eq.~\eqref{eq:hp} denote the volume average
of the scalar product $\bm{A}_p\bm{B}_p$.
One can show that the time derivative of the
helicity density $h_p$ is related to the (Abelian)
Chern-Simons density of the hypercharge gauge field,
\begin{align}
\label{eq:hpdot}
\frac{d}{dt} h_p = \frac{1}{2} \big<F_{\mu\nu}\tilde{F}^{\mu\nu}\big>
= - 2\big<\bm{E}\bm{B}\big> \,.
\end{align}
According to the standard model chiral anomaly,
the Chern-Simons density $F_{\mu\nu}\tilde{F}^{\mu\nu}$ contributes in turn
to the divergence of the baryon and lepton number currents $J_B^\mu$ and $J_L^\mu$,
\begin{align}
\partial_\mu J_B^\mu = \partial_\mu J_L^\mu = N_g \left(
\frac{g_W^2}{16\pi^2}\, W_{\mu\nu}^a\tilde{W}_a^{\mu\nu} -
\frac{g_Y^2}{32\pi^2}\, F_{\mu\nu}\tilde{F}^{\mu\nu}\right) \,.
\end{align}
In combination with Eq.~\eqref{eq:hpdot}, the time integral of this equation results
in the relation in Eq.~\eqref{eq:anomaly}.


For our purposes, the important conclusion from Eq.~\eqref{eq:anomaly} is
that a decaying hypermagnetic helicity, $\dot{h}_p \neq 0$, induces nonzero
baryon and lepton number.
To properly track the evolution of baryon number $B$ as a function of time during this process,
one needs to solve a coupled system of kinetic equations, which take into account all
relevant effects; see~\cite{Kamada:2016eeb,Kamada:2016cnb} for details.
As it turns out, $B$ is fixed after EWSB, i.e., at $T \sim 100\,\textrm{GeV}$,
since after EWSB, baryon and lepton number are no longer anomalously violated.
In the kinetic equations,
the generation of baryon number because of the time-dependent hypermagnetic helicity
is characterized by a temperature-dependent source term, $S = f\mathcal{S}$, which 
factorizes into two contributions, 
\begin{align}
f\left(\theta_W,T\right) = -T\,\frac{d\theta_W}{dT} \sin\left(2\theta_W\right) \,, \quad
\mathcal{S}\left(T\right) = \frac{H}{sT} \frac{h_p}{8\pi^2} \,.
\end{align}
Similarly as in the case of $h_c$ (see Eq.~\eqref{eq:ICrel2}),
we can estimate the magnitude of $h_p$ as follows,
\begin{align}
\label{eq:source}
h_p \sim \frac{\lambda_p}{2\pi} B_p^2 \quad\Rightarrow\quad
\mathcal{S} \sim \frac{H}{sT} \frac{\lambda_p B_p^2}{16\pi^3} \,,
\end{align}
where $\lambda_p$ and $B_p$ at $T \sim 100\,\textrm{GeV}$ are given in
Eqs.~\eqref{eq:BpIC} and \eqref{eq:lpIC}.%
\footnote{Here we consider the case
where the hypermagnetic field enters the inverse cascade regime prior to EWSB.}
Note that ${\mathcal S}$ is proportional to the amplitude of the
hypermagnetic helicity, $h_p$.
Meanwhile, $f$ is a function of the weak mixing angle $\theta_W$, which varies as a function
of temperature during the electroweak crossover.
For $d \theta_W/d T = 0$, the hypermagnetic helicity does not decay
and hence the source $S$ vanishes.


The production of baryon number because of the change in the weak mixing angle
has to compete with the usual washout processes because of electroweak sphalerons.
The effect of sphaleron washout is conveniently accounted for
in the kinetic equations by a transport coefficient
$\gamma_{\rm w, sph}$.
For a Higgs mass of $125\,\textrm{GeV}$, lattice simulations of the electroweak
crossover yield~\cite{DOnofrio:2014rug},
\begin{align}
\label{eq:washout}
\gamma_{\rm w, sph}
\simeq \exp\left[-147.7 + 107.9 \left(\frac{T}{130\,\textrm{GeV}}\right)\right] \,, \quad 
\textrm{for~} T \lesssim 161 \,\textrm{GeV} \,,
\end{align}
where $T \simeq 130\,\textrm{GeV}$ is just the temperature at which the
electroweak sphalerons freeze out.
The resulting kinetic equations are quite complicated and need to be solved numerically.
However, as shown in~\cite{Kamada:2016cnb}, the final baryon asymmetry is nicely reproduced
by the following compact analytical expression,
\begin{align}
\label{eq:asymmetry}
\eta_B = \frac{n_B}{s} \simeq \frac{17}{37} \left[\left(g_W^2 + g_Y^2\right)
\frac{f\left(\theta_W,T\right)\mathcal{S}}{\gamma_{\rm w, sph}}
\right]_{T = T_{\rm BAU}} \,, \quad
T_{\rm BAU} = 135\,\textrm{GeV} \,.
\end{align}
Here, the baryogenesis temperature $T_{\rm BAU}$ is chosen, so as to optimize
the agreement between the analytical result and the outcome of the numerical calculation.


\begin{figure}
\begin{center}
\includegraphics[width=0.48\textwidth]{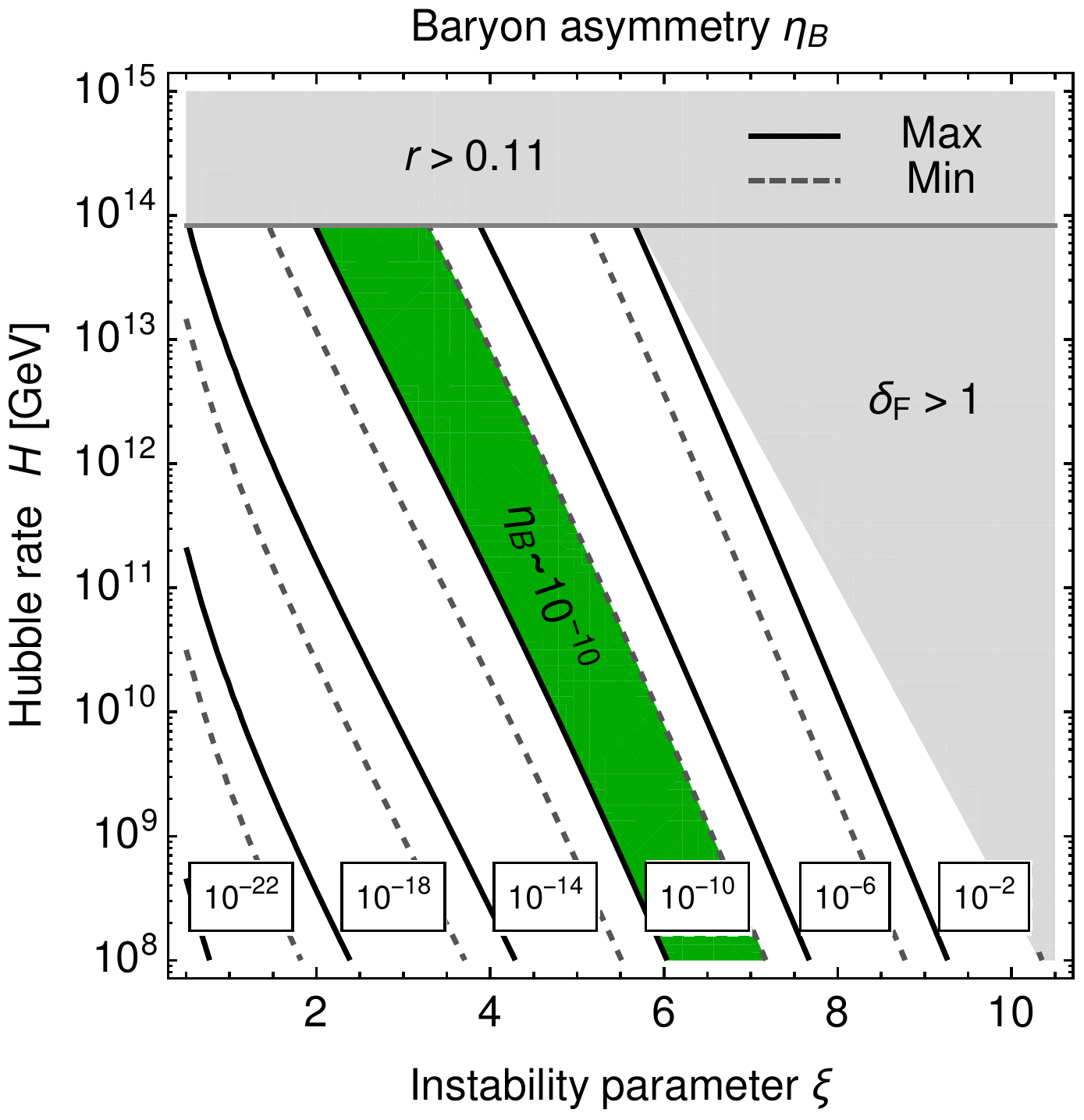}
\caption{Baryon asymmetry $\eta_B = n_B/s$ as a function of the instability
parameter $\xi$ and the Hubble rate $H$; see Eq.~\eqref{eq:etaB}.
Here, both $\xi$ and $H$ are understood to correspond to the respective
values at the end of inflation, $\xi \equiv \xi_{\rm rh}$ and $H \equiv H_{\rm rh}$.
The black solid [gray dashed] contours correspond to the maximally [minimally]
allowed value of the function $f$; see Eq.~\eqref{eq:fminmax}.
The green band illustrates the region in parameter space where $\eta_B$ is
in accord with the observed value, $\eta_B^{\rm obs} \sim 10^{-10}$.
The gray-shaded regions are the same as in Fig.~\ref{fig:deltas}.}
\label{fig:etaB}
\end{center}
\end{figure}


Combining Eqs.~\eqref{eq:BpIC}, \eqref{eq:lpIC}, \eqref{eq:source}, \eqref{eq:washout},
and \eqref{eq:asymmetry} and using $g_W \simeq 0.64$ and $g_Y \simeq 0.35$ at the electroweak
scale, we obtain the following expression for the final baryon asymmetry,
\begin{align}
\eta_B \simeq 2.9 \times 10^{-3}\,\mathcal{I}_\lambda \left[
\frac{f\left(\theta_W,T\right)}{\gamma_{\rm w, sph}}
\left(\frac{e^{2\pi\xi_{\rm rh}}}{\xi_{\rm rh}^4}\right)
\left(\frac{H_{\rm rh}^3 T^2}{M_{\rm Pl}^5}\right)^{1/2}\right]_{T = T_{\rm BAU}} \,.
\end{align}
A reliable determination of the final baryon asymmetry requires a precise
understanding of the function $f$, i.e., of the temperature dependence of
the weak mixing angle.
The latest lattice studies of the electroweak crossover, however, have a
relatively large uncertainty, as far as the exact evolution of
$\theta_W\left(T\right)$ is concerned~\cite{DOnofrio:2015gop}.
Moreover, there is a relatively large discrepancy between the numerical results
and the one-loop perturbative analytical estimate~\cite{Kajantie:1996qd}.
For this reason, we shall follow~\cite{Kamada:2016cnb} and
simply model $\theta_W$ in terms of a smooth step function,
\begin{align}
\cos^2\theta_W = \cos^2\theta_W^0 + \frac{1-\cos^2\theta_W^0}{2}
\left[1 + \tanh\left(\frac{T-T_{\rm step}}{\Delta T}\right)\right] , \:\:
\cos^2\theta_W^0 = \frac{g_W^2}{g_W^2 + g_Y^2} \simeq 0.77 \,,
\end{align}
which we believe to cover all realistic values of the function $f$ including its uncertainties.
Our phenomenological ansatz reflects the fact that, at $T \sim T_{\rm step}$,
the weak mixing angle changes from its high-temperature value in the symmetric phase,
$\cos^2\theta_W = 1$, to its low-temperature value in the Higgs phase,
$\cos^2\theta_W = \cos^2\theta_W^0$.
The width of this transition in temperature space is characterized by the parameter $\Delta T$.
Realistic values of $T_{\rm step}$ and $\Delta T$ fall into the ranges
$155\,\textrm{GeV} \lesssim T_{\rm step} \lesssim 160\,\textrm{GeV}$ and
$5\,\textrm{GeV} \lesssim \Delta T \lesssim 20\,\textrm{GeV}$, respectively.
Varying $T_{\rm step}$ and $\Delta T$ within these ranges, we find that the realistic
values of $f$ almost span three orders of magnitude,
\begin{align}
\label{eq:fminmax}
5.6 \times 10^{-4} \lesssim
f\left(\theta_W,T_{\rm BAU}\right) \lesssim 0.32 \,,
\end{align}
which translates into an uncertainty in the final baryon asymmetry,
\begin{align}
\label{eq:etaB}
\eta_B \simeq \left(1.9 \times 10^{-3} \cdots 1.1\right) \times 10^{-16}
\left(\frac{e^{2\pi\xi_{\rm rh}}}{\xi_{\rm rh}^4}\right)
\left(\frac{H_{\rm rh}}{10^{13}\,\textrm{GeV}}\right)^{3/2} \,.
\end{align}
This expression for $\eta_B$ is one of the main results of our paper.
We show $\eta_B$ as a function of $H_{\rm rh}$ and $\xi_{\rm rh}$ in Fig.~\ref{fig:etaB}.
Evidently, the observed baryon asymmetry,
$\eta_B^{\rm obs} \sim 10^{-10}$~\cite{Ade:2015xua},
can be reproduced in a large part of parameter space.
In view of Fig.~\ref{fig:etaB}, several comments are in order:


(i) For most values of the Hubble rate at the end of inflation, $H_{\rm rh}$,
the instability parameter $\xi_{\rm rh}$ needs to take a value in the range
$4 \lesssim \xi_{\rm rh} \lesssim 6$ to allow for successful baryogenesis.
According to Eq.~\eqref{eq:xiRH}, this requires the suppression scale
$\Lambda$ to take a value in the following interval,
\begin{align}
\label{eq:LambdaPred}
2.9 \times 10^{17}\,\textrm{GeV} \lesssim \Lambda \lesssim 4.3 \times 10^{17}\,\textrm{GeV} \,.
\end{align}
In other words, the requirement of successful baryogenesis roughly fixes the
value of the suppression scale $\Lambda$ in the axion-gauge-field coupling,
$\Lambda \sim 3 \times 10^{17}\,\textrm{GeV}$.
This is within a factor of $10$ of the Planck scale, which indicates
that the axion needs to be coupled rather weakly.


(ii) With $\Lambda \sim 3 \times 10^{17}\,\textrm{GeV}$ and given the location of
the green band in Fig.~\ref{fig:etaB}, it is clear that successful baryogenesis is
incompatible with large values of $\delta_{\rm F}$ and $\delta_{\rm KG}$;
see Fig.~\ref{fig:deltas} and Eq.~\eqref{eq:deltaFit}.
This means that, in the case of successful baryogenesis, the gauge field production
during inflation is never going to dominate the inflationary dynamics.
Conversely, this can be rephrased by saying that inflationary scenarios
that eventually do lead to $\delta_{\rm F} \sim 1$ unavoidably result in
an overproduction of baryon number.%
\footnote{This conclusion can be avoided if the reheating temperature
after inflation is below the electroweak scale, $T_{\rm rh} \lesssim 100\,\textrm{GeV}$.
In this case, baryon number is not anomalously violated after inflation
and the decaying \textit{magnetic} (not hypermagnetic) helicity fails to
generate a nonzero baryon asymmetry.
Similarly, our conclusions regarding the overproduction of
baryon number may change if the dynamics of reheating, which
we did not account for in our analysis, should dramatically
change our estimate of the initial hypermagnetic field strength in Eq.~\eqref{eq:initialcond}.}
Of course, this problem can be trivially solved by re-interpreting the axion
coupling to the hypercharge gauge fields as a coupling to the gauge
fields of some other, hidden $U(1)$.
But this solution comes at a high cost:
If we replaced $U(1)_Y$ by some hidden $U(1)'$, we would have to give up on
primordial magnetogenesis and baryogenesis via decaying hypermagnetic helicity
as well.
That is, we might still be able to generate a sizable signal in GWs (see Sec.~\ref{subsec:GWs});
but we would loose all other virtues of our scenario.


(iii) The parameter region consistent with successful baryogenesis is also marked
in Fig.~\ref{fig:Bp0}.
As can be seen from this figure, successful baryogenesis around the time of
EWSB correlates with a particular strength of the large-scale magnetic
fields in the present epoch, $B_p^0 \sim 10^{-16}\,\textrm{GeV}$.
Note that this is the value that we already anticipated in Eq.~\eqref{eq:benchmark}.


(iv) Our result in Fig.~\ref{fig:etaB} presents an update of earlier studies in the
literature~\cite{Anber:2015yca,Cado:2016kdp}.
In comparison to these earlier works, we find that successful baryogenesis apparently
requires larger values of $H_{\rm rh}$ as well as larger values of $\xi_{\rm rh}$.
Otherwise, the produced asymmetry will fall short off the observed value by several
orders of magnitude.
The reason for this change in numbers is that we indirectly include several effects
in our analysis that had previously been neglected.
By employing the analytical expression in Eq.~\eqref{eq:asymmetry}, we make sure
to account for the gradual change of the weak mixing angle during the electroweak crossover,
the chiral magnetic effect, the standard model Yukawa interactions, etc.
The combination of Eq.~\eqref{eq:asymmetry} with our results for $B_p$ and $\lambda_p$ at
the time of EWSB (see Eqs.~\eqref{eq:BpIC} and \eqref{eq:lpIC}) then enables us to
assess the efficiency of baryogenesis more accurately.
On the other hand, it must not be forgotten that also our analysis
still suffers from quite large uncertainties.
Future work needs to tackle in particular two issues: a better treatment
of reheating after inflation as well as a better understanding of the evolution
of the weak mixing angle during the electroweak phase transition.
Moreover, to relate the efficiency of baryogenesis to the strength
of the present-day intergalactic magnetic fields more precisely, more work
on the evolution of magnetic fields at low temperature is needed.


\subsection{High-frequency signal in gravitational waves}
\label{subsec:GWs}


In Sec.~\ref{subsec:EOMs}, we discussed the equations of motion
for the homogeneous background fields $a$ and $A_\pm^k$ in an exact FLRW
background; see Eqs.~\eqref{eq:Friedmann}, \eqref{eq:KleinGordon} and \eqref{eq:modeequation}.
In addition to this, it is also essential to study the dynamics of
the corresponding perturbations in the inflaton
field as well as in the metric tensor.
Here, a crucial observation is that the exponentially enhanced gauge field readily
provides new source terms for the primordial scalar and tensor
perturbations~\cite{Cook:2011hg,Anber:2012du}.
As it turns out, the new contributions to the scalar power spectrum are
mostly controlled by the backreaction parameter $\delta_{\rm KG}$;
see, e.g., \cite{Linde:2012bt}.
As long as we stay in the weak field regime,
$\delta_{\rm KG} \ll 1$ (see Eq.~\eqref{eq:deltaFit}), the corrections
to the scalar power spectrum are, therefore, more or less negligible
for our purposes.
The corrections to the tensor power spectrum, on the other hand, can become
quite sizable from the point of view of observational prospects\,---\,and that even so
in the weak field regime!
In fact, primordial tensor perturbations from the epoch of inflation
give rise to a spectrum of stochastic GWs in the present epoch
over a broad range of frequencies.
The amplification of the tensor power spectrum in models of
pseudoscalar inflation, therefore, has important consequences
for the expected signal of stochastic GWs from inflation.
As shown in \cite{Domcke:2016bkh, Garcia-Bellido:2016dkw},
pseudoscalar inflation may even result in sizable GWs on small scales
that are possibly within the reach of direct GW observations.
As we will discuss in the following, the primordial GW signal on
small scales ends up being dominated by the gauge contribution
rather than the vacuum contribution in a large part of parameter space.
This opens up the possibility to test our scenario, at least in principle,
by means of future GW observations.


Let us now discuss the spectral GW energy density from inflation,
$\Omega_{\rm GW}^0 h^2$, in more detail.
We first argue that the GW spectrum is flat to first approximation.
In the presence of the axion-gauge-field coupling in Eq.~\eqref{eq:aFF},
the spectral energy density $\Omega_{\rm GW}^0 h^2$ receives two contributions,
\begin{align}
\label{eq:OGW}
\Omega_{\rm GW}^0 h^2 = \left[\Omega_{\rm GW}^0 h^2\right]_{\rm vacuum} +
\left[\Omega_{\rm GW}^0 h^2\right]_{\rm gauge} \,.
\end{align}
Here, $\left[\Omega_{\rm GW}^0 h^2\right]_{\rm vacuum}$ denotes the
vacuum contribution in standard single-field slow-roll inflation,
\begin{align}
\label{eq:OGWvac}
\left[\Omega_{\rm GW}^0 h^2\right]_{\rm vacuum} =
\frac{\Omega_{\rm rad}^0 h^2}{12}
\left(\frac{g_*}{g_*^0}\right)\bigg(\frac{g_{*,s}^0}{g_{*,s}}\bigg)^{4/3}
\left(\frac{H}{\pi\,M_{\rm Pl}}\right)^2 \,,
\end{align}
which scales with the square of Hubble rate during inflation.
$\Omega_{\rm rad}^0 h^2 \simeq 2.5 \times 10^{-5}$
is the density parameter of radiation in the present epoch,
while the combination of effective numbers of DOFs ($g_* = 106.75$, $g_*^0 = 2$,
$g_{*,s} = 106.75$, $g_{*,s}^0 \simeq 3.91$)
accounts for the redshift behavior of the GW signal since its production.%
\footnote{More precisely, these factors are part of the so-called transfer
function, which describes the redshift behavior of GW modes outside and inside
the Hubble horizon; see, e.g., \cite{Buchmuller:2013lra} and references therein.
Strictly speaking, the functional form of Eq.~\eqref{eq:OGWvac} only applies to
those modes which cross inside the Hubble horizon prior to matter-radiation equality.
This is however the case for all GW modes that we are going to be interested in.}
For typical values of $H$, Eq.~\eqref{eq:OGWvac} yields a rather weak GW signal,
\begin{align}
\label{eq:OGWvac0}
\left[\Omega_{\rm GW}^0 h^2\right]_{\rm vacuum} \simeq 2.3 \times 10^{-22}
\left(\frac{H}{10^{11}\,\textrm{GeV}}\right)^2 \,.
\end{align}
Recalling that the Hubble rate during inflation is related to the primordial
tensor-to-scalar ratio,
$H \simeq 7.9 \times 10^{13}\,\textrm{GeV} \left(r/0.1\right)^{1/2}$, we point out
that Eq.~\eqref{eq:OGWvac0} is in fact equivalent to Eq.~\eqref{eq:OGWr}.


Meanwhile, one obtains for the contribution
to $\Omega_{\rm GW}^0 h^2$~\cite{Cook:2011hg,Anber:2012du} from the gauge fields,
\begin{align}
\label{eq:OGWgauge}
\left[\Omega_{\rm GW}^0 h^2\right]_{\rm gauge} \simeq
\left[\Omega_{\rm GW}^0 h^2\right]_{\rm vacuum}
\left(\frac{H}{M_{\rm Pl}}\right)^2
\left(f_L + f_R\right) e^{4\pi\xi} \,,
\end{align}
where $f_L$ and $f_R$ are two fit functions
that need to be determined numerically,%
\footnote{In Sec.~\ref{subsec:backreaction}, we solved all relevant
momentum integrals by ourselves; see Eq.~\eqref{eq:integrals}.
However, in our discussion of the primordial tensor perturbations, we will
now rely on the numerical fit functions available in the literature.}
\begin{align}
f_L = 10^{-7} \times
\begin{cases}
2.6 \> /\> \xi^{5.7} & ; \quad \xi \lesssim 3 \\
4.3 \> /\> \xi^{6.0} & ; \quad \xi \gtrsim 3
\end{cases} \,, \qquad
f_R = \frac{9.2}{\xi^{6.0}} \times 10^{-10} \,.
\end{align}
Fitting Eq.~\eqref{eq:OGWgauge}
as a function of $H^4$ and $e^{4\pi\xi}$ results in the following
phenomenological expression, which reproduces the exact result very
accurately in the entire parameter space of interest,
\begin{align}
\label{eq:OGWgauge0}
\left[\Omega_{\rm GW}^0 h^2\right]_{\rm gauge} \simeq 2.3 \times 10^{-22}\,
\exp\left[0.91 \times 4\pi\left(\xi-4.61\right)\right]
\left(\frac{H}{10^{11}\,\textrm{GeV}}\right)^4 \,.
\end{align}
Note that, in Eqs.~\eqref{eq:OGWvac0} and \eqref{eq:OGWgauge0}, we have chosen
the reference values for $H$ and $\xi$ such that both contributions to the GW
spectrum are of the same size.
Moreover, Eqs.~\eqref{eq:OGWvac0} and \eqref{eq:OGWgauge0} also illustrate
that, for $H = 10^{11}\,\textrm{GeV}$ and $\xi > 4.61$, the gauge contribution
to $\Omega_{\rm GW}^0 h^2$ exceeds the vacuum contribution.
This demonstrates that the GW signal can indeed be dominated by the gauge
contribution, although both backreaction parameters, $\delta_{\rm F}$ and $\delta_{\rm KG}$,
actually take small values; see Eq.~\eqref{eq:deltaDef}.
In fact, it is easy to show that the GW spectrum is always dominated by the gauge contribution
as soon as $H$ is larger than some critical, $\xi$-dependent value $H_{\rm GW}^{\rm crit}$,
\begin{align}
\label{eq:HGWcrit}
H_{\rm GW}^{\rm crit} = \left(f_L + f_R\right)^{-1/2}e^{-2\pi\xi}\, M_{\rm Pl}
\simeq 1.1 \times 10^{10}\,\textrm{GeV}\,\exp\left[-0.88 \times 2\pi\left(\xi-5\right)\right]\,.
\end{align}
As long as $\xi$ and $H$ are constant,
both $\left[\Omega_{\rm GW}^0 h^2\right]_{\rm vacuum}$ and 
$\left[\Omega_{\rm GW}^0 h^2\right]_{\rm gauge}$ are independent of time $t$ and frequency $f$.
In this limit, GWs therefore exhibit a flat power spectrum.


Next, let us discuss the frequency dependence of this spectrum.
We just saw that the GW spectrum is flat in the limit where $\xi$ and $H$ are constant.
However, $\xi$ and $H$ are not \textit{exactly} constant but slowly vary during inflation.
This results in a frequency dependence of the GW spectrum, after all.
A GW signal at frequency $f$ corresponds to a primordial tensor perturbation
with wavenumber $k = 2\pi R_0 f$, where $R_0$ denotes the present-day
value of the scale factor.
During inflation, the amplitude of this perturbation mode freezes out
once it is sufficiently far outside the Hubble horizon, if there
are no active sources on \textit{super}-horizon scales.
In standard slow-roll inflation without any additional coupling
to gauge fields, this requirement is satisfied simply once the $k$ mode
exits the horizon at $k/R\left(t_k\right) = H\left(t_k\right) $
(where $t_k$ is defined by this very relation).
In this case, one finds the GW amplitude at frequency $f$ by evaluating
the spectral energy density
$\Omega_{\rm GW}h^2$ for $R\left(t_k\right)H\left(t_k\right)  = k = 2\pi R_0 f$.
On the other hand, it is not \textit{a priori} clear whether this statement also remains
true if the inflaton couples to gauge fields.
The axion-gauge-field coupling may, e.g., affect the evolution of the tensor
modes even on \textit{super}-horizon scales.
However, for $\xi \sim \mathcal{O}\left(1\right)$, it turns out that the $k$ mode of
the gauge field as well as the tensor perturbations of the metric
are amplified only around the time of horizon exit.
We therefore conclude that the GW spectrum at wavenumber $k$
is generated and fixed once the $k$ mode exits the horizon.
For this reason, we can simply evaluate $\xi$ and $H$ in Eq.~\eqref{eq:OGWgauge}
at the time of horizon exit,
\begin{align}
\xi = \xi\left(t_k\right) \,, \quad H = H \left(t_k\right) \,, \quad\
R\left(t_k\right)H\left(t_k\right)  = k = 2\pi R_0 f \,.
\end{align}
Since both $\xi$ and $H$ slightly vary with time during inflation,
this procedure results in a \textit{frequency-dependent} spectrum of stochastic GWs.
The contribution from the gauge fields has an exponential dependence on $\xi$,
which results in a peak in the GW spectrum when $\xi$ is maximal.


The present frequency $f$ of the GW mode with wavenumber $k = 2\pi R_0 f$
is related to the number of $e$-folds between the time of horizon exit
and the end of inflation, $N_e$, as follows,
\begin{align}
N_e\left(f\right) = \ln\left[\frac{1}{2\pi f}\left(\frac{\pi^2}{45}
\frac{g_*\,g_{*,s}^0}{g_{*,s}}\right)^{1/3}
\frac{T_0\, T_{\rm rh}^{1/3}H_{\rm inf}^{1/3}}{M_{\rm Pl}^{2/3}}\right] \,,
\end{align}
where $H_{\rm inf} \approx H_{\rm rh}$ is the Hubble rate during inflation.
In the approximation of instant reheating, $T_{\rm rh} = \sqrt{M_* H_{\rm rh}}$,
this expression reduces to
\begin{align}
\label{eq:Nef}
N_e\left(f\right) \simeq 2.0 + \frac{1}{2} \ln\left(\frac{H_{\rm rh}}{10^{11}\,\textrm{GeV}}\right)
- \ln\left(\frac{f}{1\,\textrm{MHz}}\right) \,.
\end{align}
Then, we obtain the frequency-dependent GW spectrum as the following expression,
\begin{align}
\Omega_{\rm GW}h^2\left(f\right) =
\Omega_{\rm GW}h^2\left(\xi\left(N_e\left(f\right)\right),H\left(N_e\left(f\right)\right)\right) \,.
\end{align}


\begin{figure}
\begin{center}
\includegraphics[width=0.48\textwidth]{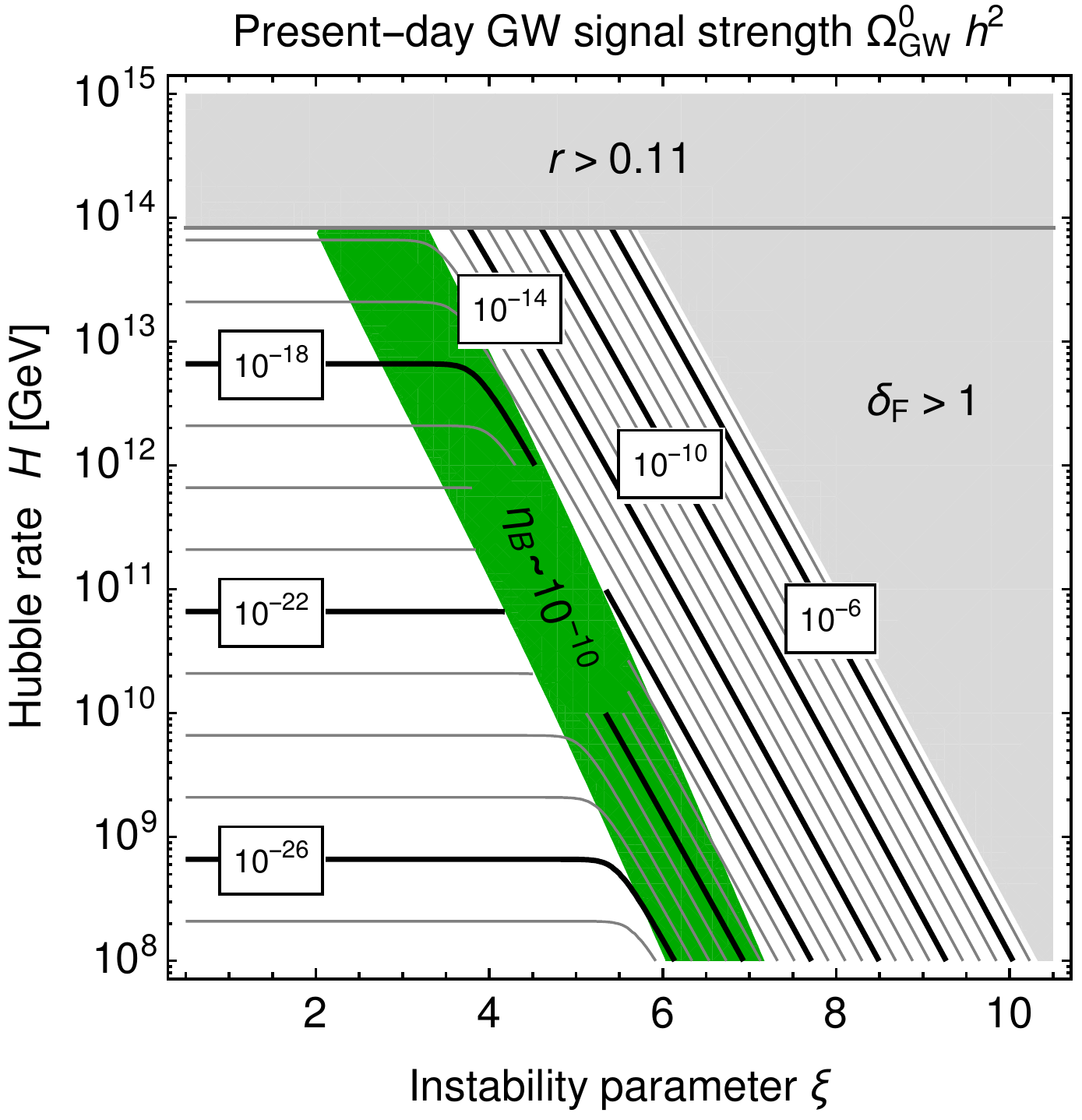}
\caption{Present-day GW signal strength $\Omega_{\rm GW}^0 h^2$
as a function of the instability parameter $\xi$ and the Hubble rate $H$;
see Eqs.~\eqref{eq:OGW}, \eqref{eq:OGWvac}, and \eqref{eq:OGWgauge}.
Here, $\xi$ and $H$ correspond to free parameters, which vary in the course of inflation.
The green band is the same as in Figs.~\ref{fig:Bp0} and \ref{fig:etaB}.
The gray-shaded regions are the same as in Fig.~\ref{fig:deltas}.}
\label{fig:OGW}
\end{center}
\end{figure}


The interplay between both contributions to the GW spectrum
is depicted in Fig.~\ref{fig:OGW}, where we plot the total
spectral energy density $\Omega_{\rm GW}^0 h^2$ as a function of $\xi$ and $H$
that slightly vary during inflation.%
\footnote{During slow-roll inflation, $\xi$ and $H$ vary
only very slowly, such that their time dependence does not
have a strong impact on the gauge field evolution.
This is the reason why we are able to solve Eq.~\eqref{eq:Whittaker} for constant $\xi$.}
Each inflation model defines a trajectory $\gamma$ in the $\xi$--$H$
plane that may, e.g., be parametrized in terms of the number of e-folds until
the end of inflation,
\begin{align}
\gamma = \left\{\left(\xi\left(N_e\right),H\left(N_e\right)\right) \forall\: N_e \right\} \,.
\end{align}
$\gamma$ passes through various values of
$\Omega_{\rm GW}^0 h^2$ during inflation.
For each model, this results in a characteristic spectrum
of stochastic GWs that could, in principle, be still observed today.


For many models of pseudoscalar inflation, $\xi$ grows towards the end of inflation, as
the inflaton velocity $\dot{a}$ becomes larger and larger.
If this growth in $\xi$ is strong enough, such that $H > H_{\rm GW}^{\rm crit}$
at some point (see Eq.~\eqref{eq:HGWcrit}), the gauge contributions to
$\Omega_{\rm GW}h^2$ will result in an exponentially steep increase in the GW spectrum.
This mechanism of GW production will shut off as soon as inflation is over
and our mechanism of gauge field production is no longer active.
All in all, we therefore expect a characteristic feature in the GW spectrum associated
with the explosive gauge field production at the end of inflation, i.e., around $N_e \simeq 0$.
According to Eq.~\eqref{eq:Nef}, we estimate that this peak should occur
at frequencies in the MHz range or at even higher frequencies,
\begin{align}
\label{eq:fpeak}
N_e\left(f_{\rm peak}\right) \simeq 0 \quad\Rightarrow\quad
f_{\rm peak} \simeq 7.1 \,\textrm{MHz}
\left(\frac{H_{\rm rh}}{10^{11}\,\textrm{GeV}}\right)^{1/2} \,.
\end{align}


To estimate the strength of the peak in the GW spectrum, we simply need
to evaluate $\Omega_{\rm GW}h^2$ in Eq.~\eqref{eq:OGW}
for $\xi = \xi_{\rm rh}$ and $H = H_{\rm rh}$.
Or alternatively, we may trade the dependence on $\xi_{\rm rh}$ and $H_{\rm rh}$
for the present-day strength of the magnetic field, $B_p^0$, as well as the peak
frequency, $f_{\rm peak}$.
Making use of Eqs.~\eqref{eq:Bp0}, \eqref{eq:OGW}, and \eqref{eq:fpeak},
we then find the following numerical relation,
\begin{align}
\label{eq:OGWBf}
\left[\Omega_{\rm GW}^0 h^2\right]_{\rm peak} \simeq 3.2\times 10^{-20} \:
\bigg(\frac{B_p^0}{10^{-16}\,\textrm{GeV}}\bigg)^{6.13}
\bigg(\frac{f_{\rm peak}}{10\,\textrm{MHz}}\bigg)^{1.87} \,,
\end{align}
which is another main result of our paper.
In order to eliminate the $\xi$ dependence in $\Omega_{\rm GW}h^2$,
we numerically solved Eq.~\eqref{eq:Bp0} for $\xi$.
Based on the relation in Eq.~\eqref{eq:OGWBf},
we plot $B_p^0$ as a function of $f_{\rm peak}$
and $\left[\Omega_{\rm GW}^0 h^2\right]_{\rm peak}$ in Fig.~\ref{fig:OGWf}.
In view of Eq.~\eqref{eq:OGWBf} and Fig.~\ref{fig:OGWf},
several comments are in order:


\begin{figure}
\begin{center}
\includegraphics[width=0.48\textwidth]{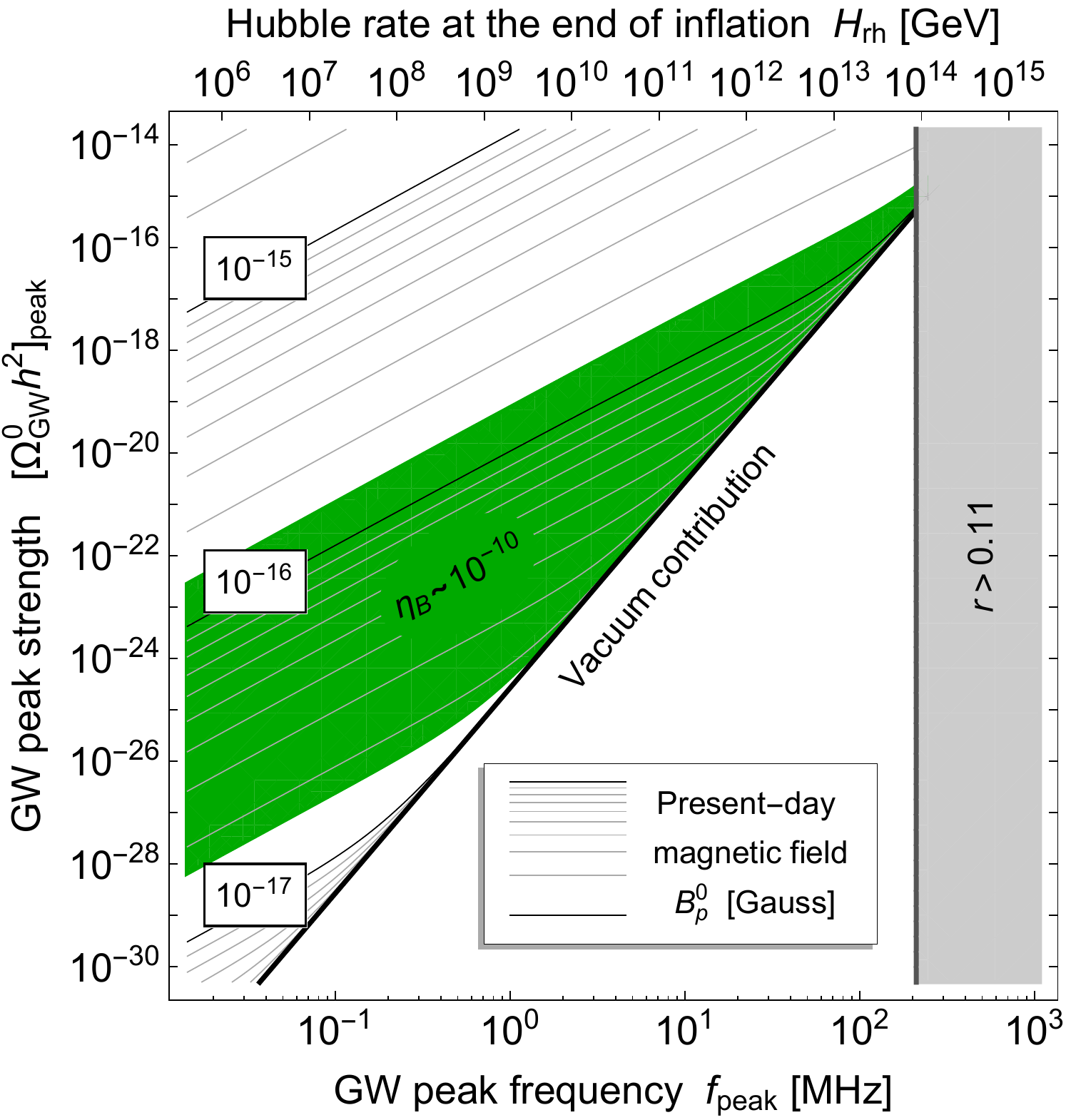}
\caption{Present-day magnetic field strength $B_p^0$
as a function of the peak frequency $f_{\rm peak}$ and
the strength of the peak in the GW spectrum associated
with the gauge field production at the end of inflation,
$\left[\Omega_{\rm GW}^0 h^2\right]_{\rm peak}$;
see Eq.~\eqref{eq:OGWBf}.
In the approximation of instant reheating, $f_{\rm peak}$ is directly related
to the Hubble rate at the end of inflation; see Eq.~\eqref{eq:fpeak}.
The green band illustrates the region in parameter space where $\eta_B \sim 10^{-10}$;
see Eq.~\eqref{eq:etaB}.}

\label{fig:OGWf}
\end{center}
\end{figure}


(i) In scenarios consistent with successful baryogenesis,
the gauge field production at the end of inflation is typically accompanied
by a rather weak signal in GWs at high frequencies.
The detection of this peak is certainly out of reach of present-day technology.
On the other hand, it is an unavoidable consequence of primordial magnetogenesis
in our scenario.
In the future, the detection of such a GW peak may therefore serve as a smoking-gun signal
of primordial magnetogenesis at the end of pseudoscalar inflation.
This in turn would lend support to the idea of baryogenesis from decaying hypermagnetic helicity.
In particular, one could assess whether the strength of the observed GW
signal turns out to be consistent with an inflaton coupling to the
hypercharge gauge field\,---\,or whether this assumption would
lead to baryon overproduction.


(ii) Along the diagonal line in Fig.~\ref{fig:OGWf}, the GW spectrum at the
end of inflation is dominated by the (irreducible) vacuum contribution.
The part of parameter space below this line is therefore not accessible.
Meanwhile, the vertical distance between this line and any point above indicates
the extent to which the peak in the GW spectrum sticks out of the usual vacuum background.


(iii) We stress once more that, at the quantitative level,
Eq.~\eqref{eq:OGWBf} and Fig.~\ref{fig:OGWf} may still receive a number of corrections.
After all, every quantity in our analysis ($B_p^0$, $\eta_B$, $\Omega_{\rm GW}h^2$) comes
with potentially large uncertainties.
Nonetheless, we believe that Eq.~\eqref{eq:OGWBf} and Fig.~\ref{fig:OGWf} convey the correct
idea at the qualitative level.
Our results illustrate that pseudoscalar inflation leads to a highly nontrivial
relation between initially completely independent phenomena: the present-day strength
of the intergalactic magnetic field, the baryon asymmetry of the universe,
and the stochastic background of GWs.
This realization is one of our main achievements in this paper.


\section{Explicit scenarios based on natural inflation}
\label{sec:example}


All quantities that we were interested in so far ($B_p^0$, $\lambda_p^0$, $\eta_B$,
and $\Omega_{\rm GW}^0 h^2$) solely depend on the values of $\xi$ and $H$
at the end of inflation.
This observation allowed us to perform a completely model-independent analysis
up to this point.
We did not specify the form of the inflaton potential $V\left(a\right)$ and
discarded all details of the reheating process.
Instead, we simply employed a model-independent parametrization in terms of
$\xi_{\rm rh}$ and $H_{\rm rh}$.
This means that all of our results up to this point apply to any model of pseudoscalar
inflation that is anomalously coupled to the standard model hypercharge sector.
Now, however, we shall illustrate our results
by means of concrete examples, in order to see how realistic
models of inflation give rise to the phenomenology described in the previous
sections.
To this end, we shall now study the evolution of $\xi$ and $H$ during inflation in concrete
models and illustrate how they approach certain values towards the end of inflation.
In other words: up to now, we were only interested in certain points in
the $\xi$--$H$ parameter plane; now we turn to the
\textit{inflationary trajectories} in this parameter plane.


Given the Lagrangian in Eq.~\eqref{eq:Lagrangian}, the inflaton field $a$
is naturally identified as an axion, {\it i.e.}, the PNGB of a spontaneously broken
global symmetry $G_{\rm global}$.
If this symmetry is anomalous under the standard model hypercharge
gauge group $U(1)_Y$, the axion $a$ will couple to the standard model hypercharge
gauge field just as in Eq.~\eqref{eq:aFF}.
Moreover, if $G_{\rm global}$ is \textit{in addition} anomalous under some strongly
coupled gauge symmetry $G_{\rm strong}$, nonperturbative effects in the $G_{\rm strong}$
gauge sector will generate a scalar potential for $a$ of the following form,
\begin{align}
\label{eq:Vcosine}
V\left(a\right) = m_a^2\, f_a^2 \left[1-\cos\left(\frac{a}{f_a}\right)\right] \,.
\end{align}
Here, $m_a$ and $f_a$ denote the axion mass as well as the axion decay constant.
The overall scale of the axion potential is set by the confinement scale in
the strongly coupled sector, $m_a^2 f_a^2 \sim \Lambda_{\rm strong}^4$.
In the following, we can treat both $m_a$ and $f_a$ as free parameters.
The scalar potential in Eq.~\eqref{eq:Vcosine} is nothing but the scalar potential
of natural inflation~\cite{Freese:1990rb,Adams:1992bn}.
This is a trivial statement given the fact that natural inflation
denotes the very idea that inflation is driven by the PNGB of some
spontaneously broken and anomalous global symmetry.
In the following, we shall study the inflationary trajectory for natural inflation
in the $\xi$--$H$ plane.


For any value of the inflaton field during slow-roll inflation,
one can determine $\left(\xi,H\right)$ from Eqs.~\eqref{eq:Friedmann}
and \eqref{eq:KleinGordon}.
Once we replace $\dot{a}$ by $2\Lambda H\xi$, see Eq.~\eqref{eq:xi},
and neglect $\ddot{a}$, we have
\begin{align}
\label{eq:sec5}
3M_{\rm Pl}^{2}H^{2}-\frac{1}{2}\left(2\Lambda H\xi\right)^{2}-
V - \frac{1}{2}\>\big[\,\rho_{EE}\left(\xi,H\right)+\rho_{BB}\left(\xi,H\right)\big] & = 0
\,, \\ \nonumber
6\Lambda H^{2}\xi+\frac{dV}{da}-\frac{1}{\Lambda}\,\rho_{EB}\left(\xi,H\right) & = 0
\,.
\end{align}
For a given pair of values for $\left(V,\frac{dV}{da}\right)$ as well as
for given $\Lambda$, we can numerically solve Eq.~\eqref{eq:sec5} for $\left(\xi,H\right)$.
The slow-roll parameter $\varepsilon$
including the contribution of the gauge field is~\cite{Anber:2009ua}
\begin{align}
\label{eq:sec5-1}
\varepsilon = -\frac{\dot H}{H^2} =
\frac{1}{2M_{\rm Pl}^{2}H^{2}} \left[ \dot{a}^2 +
\frac{2}{3}\,\big(\rho_{EE}+\rho_{BB}\big)\right] \,,
\end{align}
which can be computed once $\left(\xi,H\right)$ has been determined.
For each field value, we are therefore able to compute
the corresponding value of $\varepsilon$.
With the aid of Eq.~\eqref{eq:sec5-1},
we can hence numerically determine the end point of inflation,
where $\varepsilon = 1$.
The number of e-folds $N_{e}$ is given by
\begin{align}
\label{eq:sec5-2}
N_e = \int_{a_{\rm end}}^{a}da\,\frac{dN_{e}}{da} \,,
\end{align}
where $a_{\rm end}$ is the field value at the end of inflation and
the integrand is a simple function of $\xi$,
\begin{align}
\frac{dN_{e}}{da}=-\frac{H}{\dot{a}}=-\frac{1}{2\Lambda\xi}.\label{eq:sec5-3}
\end{align}
In summary, for each inflaton field value, we can compute the quadruplet
$\left(\xi,H,\varepsilon,N_{e}\right)$, which
enables us to draw an inflationary trajectory for any given model in the
$\xi$--$H$ parameter plane.


Now let us compute some explicit examples.
Our variant of the natural inflation model is
characterized by three parameters:
the two parameters $\left(m_{a},f_{a}\right)$
in the potential, see Eq.~\eqref{eq:Vcosine}, as well as the suppression scale
$\Lambda$ in the Chern-Simons interaction.
We take the following values:%
\footnote{The parameters $f_a$ and $\Lambda$ ought to be related
to each other in the UV completion of our model.
But in this study, we do not specify any UV physics,
which is why we treat $f_a$ and $\Lambda$ as independent parameters.}
\begin{align}
\label{eq:sec5-5}
\textrm{Model A:} \qquad
m_{a}^{2} = 4.1 \times 10^{-11}\, M_{\rm Pl}^{2},\quad
f_{a} = 7.0\, M_{\rm Pl},\quad
\Lambda^{-1} = 5.6\, M_{\rm Pl}^{-1} \,.
\end{align}
Here, to distinguish it from other models that will be discussed, we
refer to it as model A.
Note that, to ensure successful baryogenesis,
$\Lambda$ cannot be chosen arbitrarily; see Eq.~\eqref{eq:LambdaPred}.
Besides, to make the model compatible with the CMB observations, we
need to chose particular values for the two parameters $m_{a}$ and $f_{a}$.
The parameters
in Eq.~(\ref{eq:sec5-5}) have been tuned in such a way that the model
is compatible with all CMB observations and the baryon number asymmetry.

\begin{figure}
\begin{center}

\includegraphics[width=7.5cm]{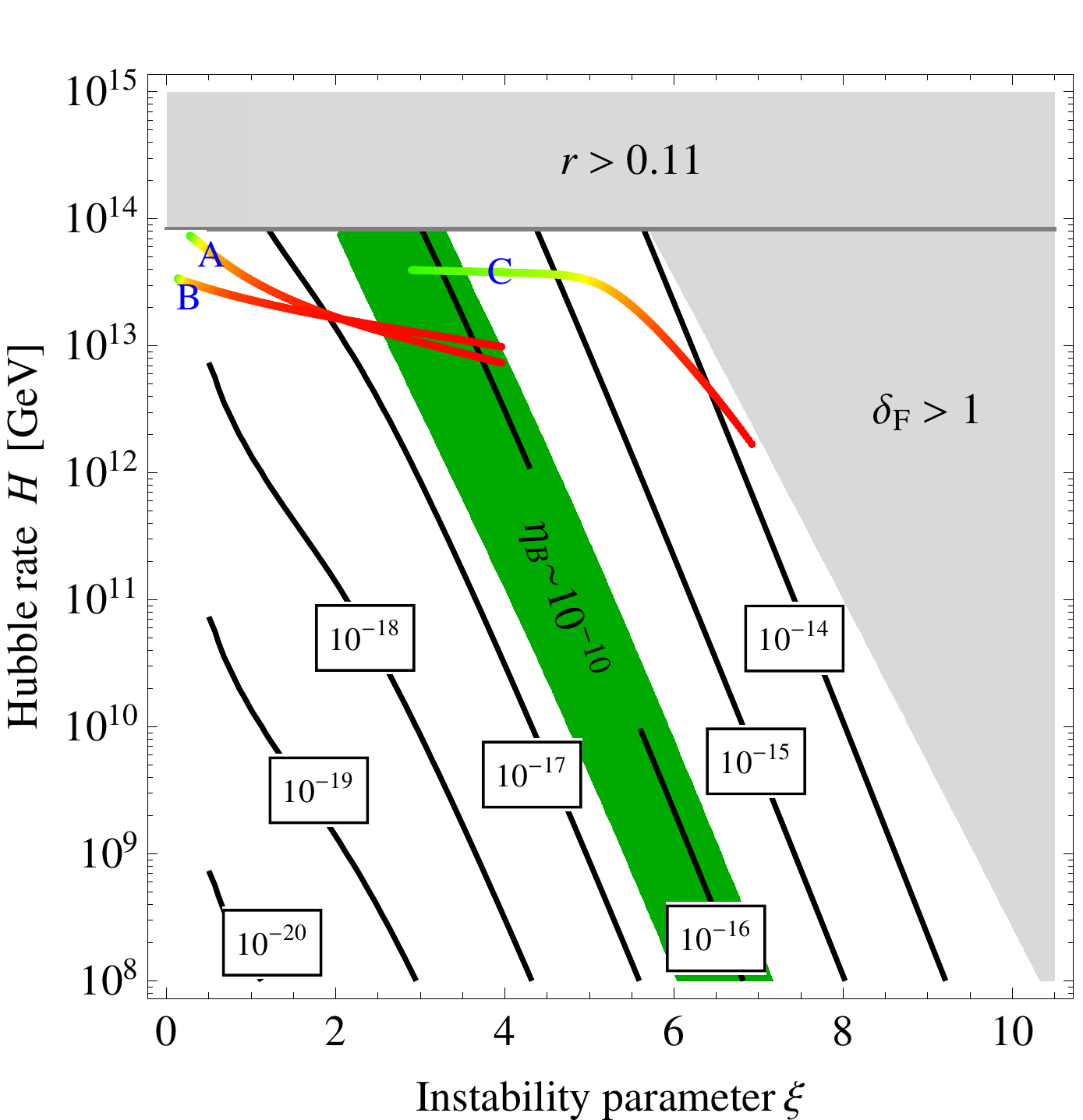}\ \ \raisebox{0.12\height}{
\includegraphics[height=6.5cm]{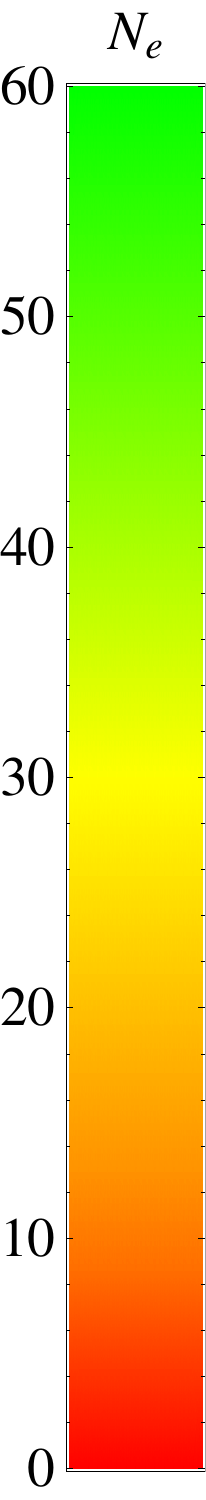}}
\caption{Trajectories of several inflation models in the $\xi$--$H$ parameter plane.
Trajectory A corresponds to natural inflation, see Eqs.~\eqref{eq:Vcosine}
and \eqref{eq:sec5-5}, while trajectories B and C correspond to Starobinsky inflation,
see Eq.~\eqref{eq:sec5-11}, in the case of small and large
axion-gauge-fields coupling, respectively.
Numerical details are listed in Tab.~\ref{tab:numeric}.
Successful baryogenesis is accomplished for any inflationary trajectory
that ends in the green band, i.e., if the point 
$(\xi,H)_\text{end}=(\xi_{\rm rh},H_{\rm rh})=\big(\xi(N_e=0),H(N_e=0)\big)$ 
lies in the green band; see Fig.~\ref{fig:Bp0} and Eq.~\eqref{eq:etaB}.}
\label{fig:traj}
\end{center}
\end{figure}

Following the procedure introduced above, we numerically compute $\left(\xi,H\right)$
by solving Eq.~(\ref{eq:sec5}) for each field value in the relevant part of the potential
with a step width of $\Delta a = 0.01\, M_{\rm Pl}$.
Then we compute the slow-roll parameter $\varepsilon$
to determine the end of inflation, which is at
\begin{align}
\label{eq:sec5-6}
\textrm{Model A:} \qquad
a_{\rm end} = -0.94\, M_{\rm Pl} \,.
\end{align}
For all field values during inflation, $ a < a_{\rm end}$, the number of e-folds $N_{e}$ is
computed according to Eq.~(\ref{eq:sec5-2}).
Together, these points form the trajectory corresponding
to model A in Fig.~\ref{fig:traj}.


\begin{table}[t]
\begin{center}
\begin{tabular}{cccc}
\hline
Model & A & B & C\tabularnewline
\hline
$\Lambda^{-1}$ & $5.6\, M_{\rm Pl}^{-1}$ & $5.6\, M_{\rm Pl}^{-1}$
& $75\, M_{\rm Pl}^{-1}$\tabularnewline\rule[0ex]{0pt}{3ex}  $V(a)$
& Eq.~(\ref{eq:Vcosine})
& Eq.~(\ref{eq:sec5-11}) & Eq.~(\ref{eq:sec5-11})\tabularnewline
\rule[0ex]{0pt}{6ex}  $\left[\begin{array}{c}
m_{a}^{2}\\
f_{a}
\end{array}\right]$ or $\left[\begin{array}{c}
V_{0}\\
\gamma_{\rm s}
\end{array}\right]$ & $\left[\begin{array}{c}
4.1\times10^{-11}\, M_{\rm Pl}^{2}\\
7.0\, M_{\rm Pl}
\end{array}\right]$ & $\left[\begin{array}{c}
6.7\times10^{-10}\, M_{\rm Pl}^{4}\\
0.30\, M_{\rm Pl}^{-1}
\end{array}\right]$ & $\left[\begin{array}{c}
1.0\times10^{-9}\, M_{\rm Pl}^{4}\\
0.30\, M_{\rm Pl}^{-1}
\end{array}\right]$\tabularnewline
\rule[0ex]{0pt}{10ex}  $\left[\begin{array}{c}
a\\
\dot{a}\\
H\\
\xi
\end{array}\right]_{N_{e}=0}$ & $\left[\begin{array}{c}
-0.94\, M_{\rm Pl}\\
4.3\times10^{-6}\, M_{\rm Pl}^{2}\\
3.0\times10^{-6}\, M_{\rm Pl}\\
4.0
\end{array}\right]$ & $\left[\begin{array}{c}
-0.83\, M_{\rm Pl}\\
5.7\times10^{-6}\, M_{\rm Pl}^{2}\\
4.0\times10^{-6}\, M_{\rm Pl}\\
4.0
\end{array}\right]$ & $\left[\begin{array}{c}
-0.09\, M_{\rm Pl}\\
1.3\times10^{-7}\, M_{\rm Pl}^{2}\\
6.9\times10^{-7}\, M_{\rm Pl}\\
6.9
\end{array}\right]$\tabularnewline
\rule[0ex]{0pt}{10ex}  $\left[\begin{array}{c}
a\\
\dot{a}\\
H\\
\xi
\end{array}\right]_{N_{e}=55}$ & $\left[\begin{array}{c}
-13\, M_{\rm Pl}\\
3.0\times10^{-6}\, M_{\rm Pl}^{2}\\
3.0\times10^{-5}\, M_{\rm Pl}\\
0.28
\end{array}\right]$ & $\left[\begin{array}{c}
-8.7\, M_{\rm Pl}\\
6.7\times10^{-7}\, M_{\rm Pl}^{2}\\
1.4\times10^{-5}\, M_{\rm Pl}\\
0.13
\end{array}\right]$ & $\left[\begin{array}{c}
-7.2\, M_{\rm Pl}\\
1.3\times10^{-6}\, M_{\rm Pl}^{2}\\
1.6\times10^{-5}\, M_{\rm Pl}\\
2.9
\end{array}\right]$\tabularnewline
\rule[-6ex]{0pt}{14ex}   $\left[\begin{array}{c}
P_s\\
n_s\\
r
\end{array}\right]_{N_{e}=55}$ & $\left[\begin{array}{c}
2.3\times10^{-9}\\
0.96\\
0.08
\end{array}\right]$ & $\left[\begin{array}{c}
2.1\times10^{-9}\\
0.97\\
0.02
\end{array}\right]$ & $\left[\begin{array}{c}
2.2\times10^{-9}\\
0.94\\
0.05
\end{array}\right]$\tabularnewline
\hline
\end{tabular}
\caption{Various parameters and numerical results for the three models A, B, and C.}
\label{tab:numeric}
\end{center}
\end{table}


The scalar power spectrum is evaluated according to
\begin{align}
\label{eq:sec5-4}
P_s = \left(\frac{H^{2}}{2\pi\dot{a}}\right)^{2}
\left(\frac{k}{k_{\rm CMB}}\right)^{n_{s}-1} \,,
\end{align}
where $k_{\rm CMB} = 0.05\,\textrm{Mpc}^{-1}$ is the CMB pivot scale;
and $H$ and ${\dot a}$ are evaluated at the time when the pivot scale exits the horizon.
We neglect the contribution from the gauge fields, since
it is negligibly small in the region of interest.
This should be compatible with the CMB normalization~\cite{Ade:2015lrj}:
\begin{align}
\label{eq:sec5-10}
P_s^{\rm obs} = \left(2.21 \pm 0.07\right) \times 10^{-9} \,.
\end{align}
The CMB pivot scale exits the horizon at $N_e^{\rm CMB} \simeq 55$,
where we obtain
\begin{align}
\textrm{Model A:} \qquad
a = - 13.4\, M_{\rm Pl} \,,\quad
\dot{a} = 3.0 \times 10^{-6}\, M_{\rm Pl}^{2} \,, \quad
H = 3.0 \times 10^{-5}\, M_{\rm Pl} \,,\quad \xi = 0.28 \,.
\end{align}
With these numerical results, we can evaluate the scalar power spectrum:
\begin{align}
\label{eq:sec5-7}
\textrm{Model A:} \qquad
P_s=2.3\times10^{-9} \,,
\end{align}
which is compatible with Eq.~(\ref{eq:sec5-10}).
Since the gauge field contribution is very weak at $N_e = N_e^{\rm CMB}$,
the spectral index $n_{s}$ and the tensor-to-scalar ratio $r$
can be evaluated in the conventional way:
\begin{align}
\textrm{Model A:} \qquad
\label{eq:sec5-8}
n_s = 1 + 2\,\eta - 6\,\varepsilon \simeq 0.96 \,, \quad
r = 16\,\varepsilon \simeq 0.08 \,.
\end{align}
which agrees with the current PLANCK constraints~\cite{Ade:2015lrj}.
All of the above numerical results are summarized in Tab.~\ref{tab:numeric}.

\begin{figure}
\begin{center}
\includegraphics[height=5.5cm]{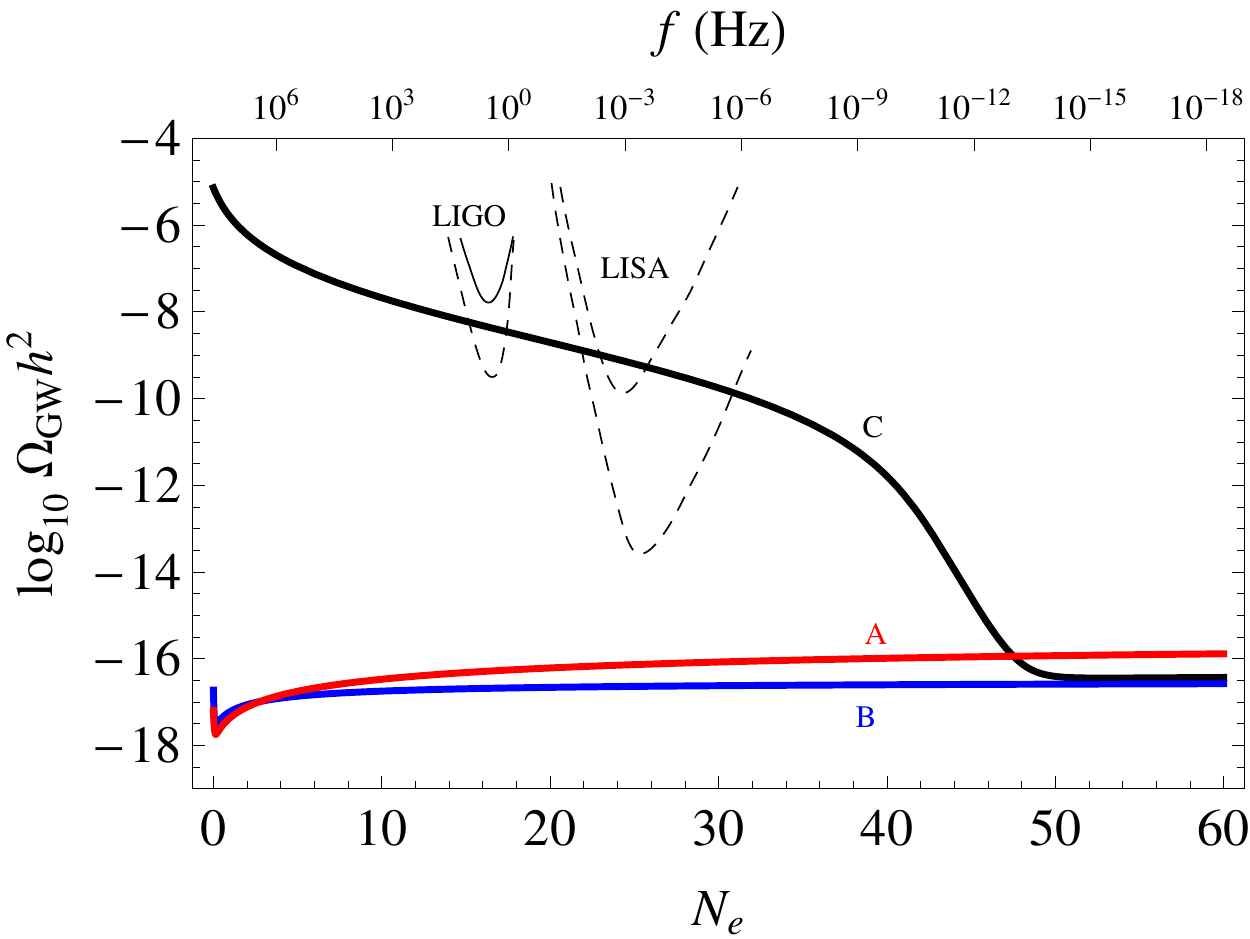}\raisebox{0.00\height}{\hspace{1cm}
\includegraphics[height=5.5cm]{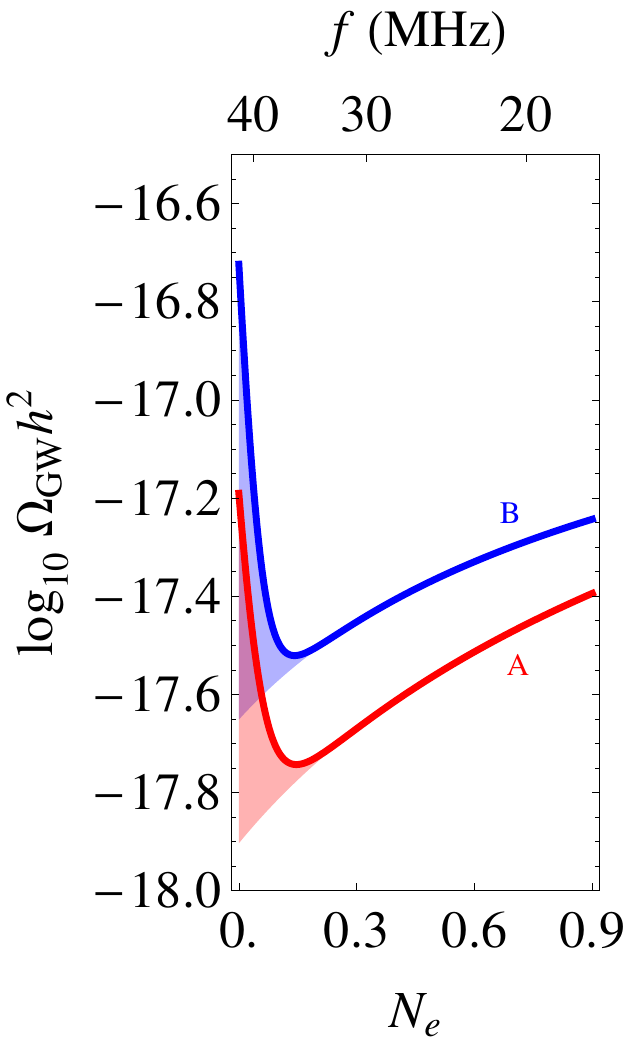}}
\caption{GW spectra of several models compared
to the current (solid lines) and future (dashed lines) constraints
from advanced LIGO and LISA.
The red, blue and black curves corresponds to models
A, B, and C, respectively.
Model A: natural inflation; model B/C: Starobinsky inflation
with a small/large axion-gauge-field coupling; see Tab.~\ref{tab:numeric}
for the numerical details.
The red and blue shadows in the right panel represent the gauge contributions.
To relate the number of e-folds $N_e$ and the frequency $f$ we have used
Eq.~\eqref{eq:Nef} with $H_{\rm rh}=10^{12.5}$ GeV, which is approximately correct
for the models A, B, and C. }
\label{fig:GW-N}
\end{center}
\end{figure}

We can further compute the GW spectrum according to
Eq.~(\ref{eq:OGW}), including both the vacuum and gauge contributions.
This is shown by the red curve in Fig.~\ref{fig:GW-N}, where the
red shadow denotes the gauge contribution.
In Fig.~\ref{fig:GW-N}, we also present the current constraints from advanced
LIGO and future sensitivities from advanced LIGO and LISA.
It turns out that the GW energy density $\Omega_{{\rm GW}}^{0}h^{2}$
produced in model A is far below the reach of current or upcoming
GW interferometers.


It has been shown \cite{Domcke:2016bkh} however that, with a strongly coupled
axion, some models could reach the sensitivity of current or upcoming GW interferometers.
For the Starobinsky model~\cite{Starobinsky:1980te}, e.g.,
\begin{align}
\label{eq:sec5-11}
V_{{\rm Starobinsky}}(a) = V_0\> \big(1-e^{\gamma_{\rm s} a}\big)^2 \,,\quad a < 0 \,,
\end{align}
with $(V_{0},\thinspace\gamma_{\rm s})=(1.0 \times 10^{-9}\, M_{\rm Pl}^{4}, 0.3\, M_{\rm Pl}^{-1})$
and an axion-gauge-field coupling $\Lambda^{-1} = 75\, M_{\rm Pl}^{-1}$ (which
is more than 10 times larger than the case we just discussed),
$\Omega_{{\rm GW}}^{0}h^{2}$ can reach the future sensitivities
of advanced LIGO and LISA, as shown by the black curve in Fig.~\ref{fig:GW-N}.
Such a large GW energy is due to a very strong axion-gauge-field coupling,
which transfers almost the entire energy carried by the inflaton into the
gauge field and induces much larger tensor perturbations.
But strong axion-gauge-field couplings will always lead
to baryon overproduction, as discussed model-independently
in Sec.~\ref{sec:BAUGWs}; see Eq.~\eqref{eq:LambdaPred}.
Indeed, Fig.~\ref{fig:traj} shows that the trajectory of this model
(we refer to it as model C; for numerical details, see Tab.~\ref{tab:numeric})
ends at a point far away from the region for successful baryogenesis
(the green band).
Actually, this point is very close to the bound $\delta_\textrm{F}=1$,
which corresponds to the situation that
the entire energy of the universe is stored in the gauge field.
If we reduce the axion-gauge-field coupling of model C to the same value
as in model A, then the model (now referred to as model B) leads to successful
baryogenesis; see Fig.~\ref{fig:traj}.
But it has small $\Omega_{{\rm GW}}^{0}h^{2}$,
approximately of the same order of magnitude as model A.


It is interesting to note that the GW spectra of model A and B
both peak at the very end of inflation, where the gauge contributions
become dominant (at $0\lesssim N_e \lesssim 0.3$,
corresponding to $f\sim 40$ MHz, see the
right panel of Fig.~\ref{fig:GW-N}).
This is an important feature of these models compared
to models without the inflaton-gauge-field coupling.
Although these peaks are out of reach of
conventional GW interferometers, once detected by some other new technology
in the future, they may serve as a smoking-gun signal for baryogenesis
via primordial magnetic fields.


\section{Conclusions}
\label{sec:conclusions}


In this paper, we revisited the implications of a Chern-Simons-like inflaton coupling
to the standard model hypercharge gauge field,
$\mathcal{L} \supset a/\left(4\Lambda\right) F\tilde{F}$, in general models
of pseudoscalar inflation.
We focused in particular on two phenomenological aspects:
(i) the production of primordial gauge fields towards the end of inflation
(i.e., primordial magnetogenesis) and its consequences
for baryogenesis from decaying (hyper)magnetic fields at the time of EWSB;
and (ii) the associated production of primordial tensor perturbations and
their impact on the present-day spectrum of stochastic gravitational waves.
Our main results can be summarized as follows:


\begin{enumerate}
\item Primordial magnetogenesis at the end of pseudoscalar inflation
can result in sizable present-day magnetic fields with a correlation
length on astrophysical scales; see Eqs.~\eqref{eq:Bp0} and \eqref{eq:lp0}.
These fields then contribute to the intergalactic magnetic fields we observe today.
The main uncertainties in our estimate are: (i) the impact of reheating on
the gauge field production after the end of inflation and (ii) the impact
of damping effects at temperatures below $10\,\textrm{MeV}$.
In particular, we point out that the presence of a strong hyper-EM field
during reheating may open up new channels of particle production, such as
pair production via the Schwinger effect.
This effect has recently been studied by the authors of Ref.~\cite{Tangarife:2017rgl},
who referred to it as \textit{Schwinger reheating}.
Moreover, it is important to understand how the emerging charged plasma
back-reacts on the primordial gauge field.
A better treatment of this complicated process requires a dedicated
numerical simulation that takes into account both nonperturbative particle
production as well as MHD effects.
Such a study is beyond the scope of this paper; but we certainly encourage
further efforts into this direction.
\item The primordial gauge fields generated towards the end of pseudoscalar
inflation are maximally helical and can, thus, source the generation of
nonzero baryon number around the time of the electroweak crossover via the
chiral anomaly in the standard model.
We updated previous studies of this mechanism of primordial baryogenesis,
which led us to the conclusion that successful baryogenesis is indeed
possible in a large part of parameter space, see Eq.~\eqref{eq:etaB}.
We found that the pseudoscalar
inflaton must be \textit{weakly} coupled to the hypercharge gauge field,
since the primordial gauge fields will otherwise result in an overproduction
of baryon number.
To be more precise, successful baryogenesis requires an instability
parameter $\xi$ of around $\xi \sim 5$ at the end of inflation, which translates
into a suppression scale $\Lambda$ of around $\Lambda \sim 3 \times 10^{17}\,\textrm{GeV}$.
Again, a main uncertainty of our estimate is the strength of the primordial
hypermagnetic field at the time of EWSB.
Besides that, the poor knowledge of the temperature dependence of the
weak mixing angle during the crossover, $\theta_W(T)$, induces further uncertainties.
A better understanding of baryogenesis via decaying helicity,
therefore, requires a more careful determination of $\theta_W(T)$.
\item The gauge field production at the end of inflation is accompanied
by the production of stochastic gravitational waves.
We are able to show that the production of gauge fields consistent with
successful baryogenesis at later times typically results a weak GW signal
at frequencies in the MHz range or even above; see Eq.~\eqref{eq:OGWBf}
and Fig.~\ref{fig:OGWf}.
GWs at such high frequencies are extremely hard to detect;
see \cite{Nishizawa:2007tn,Akutsu:2008qv} for a past measurement,
\cite{Chou:2016hbb} for an on-going experiment as well as
\cite{Cruise:2012zz,Arvanitaki:2012cn} for proposals of future techniques.
However, if the signal predicted in our scenario should eventually be measured
by future experiments, it would serve as a smoking gun for the explosive
gauge field production at the end of inflation (and hence provide
evidence for baryogenesis via decaying magnetic fields during
the electroweak crossover).
On the other hand, we are able to conclude that any stronger GW signal
would imply the overproduction of baryon number.
In this case, one would either have to give up on an inflaton coupling
to the standard model hypercharge gauge field or one would have to
assume low reheating temperature, such that $T_{\rm rh} \lesssim T_{\rm ew}$.
\end{enumerate}


Our analysis illustrates how models of pseudoscalar inflation result
in a highly non-trivial interrelation of several, \textit{a priori} unrelated
phenomena: the present-day large-scale magnetic field, the baryon asymmetry of the
universe, and features in the spectrum of stochastic GWs.
In the present paper, we mainly focused on the qualitative aspects of
this interplay of phenomena and more work is needed to arrive at
more reliable and more precise quantitative predictions.
Such an effort requires progress on several fronts.
But it also
promises to lead to a better understanding of an intriguing cosmological
scenario that comes with rich phenomenology deriving from a single
additional operator in the effective Lagrangian:
$\mathcal{L} \supset a/(4\Lambda)\>F\tilde{F}$.


\subsubsection*{Acknowledgements}


The authors would like to thank Valerie Domcke, Tomohiro Fujita, Andrew Long,
Andreas Ringwald, and Markus Rummel for valuable comments and discussions.
K.\,K.\ acknowledges support from the DOE for this work under Grant No. DE-SC0013605.
This project has received funding from the European Union's Horizon 2020
research and innovation programme under the Marie Sk\l odowska-Curie grant
agreement No.\ 674896 (K.\,S.).



\end{document}